\documentclass{article}
\usepackage{graphicx,emulateapj}

\submitted{Submitted to ApJ, Feb. 17, 200; Resubmitted, Nov. 28, 2000}

\def\myputfigure#1#2#3#4#5%
{\vskip#5pt\makebox[0pt]{\hskip#2in
\includegraphics[width=#3\textwidth]{#1}}\vskip#4pt\hfill}

\def\gsim{\;\rlap{\lower 2.5pt
 \hbox{$\sim$}}\raise 1.5pt\hbox{$>$}\;}
\def\lsim{\;\rlap{\lower 2.5pt
   \hbox{$\sim$}}\raise 1.5pt\hbox{$<$}\;}

\def\spose#1{\hbox to 0pt{#1\hss}}
\def\lta{\mathrel{\spose{\lower 3pt\hbox{$\mathchar''218$}}
     \raise 2.0pt\hbox{$\mathchar''13C$}}}
\def\gta{\mathrel{\spose{\lower 3pt\hbox{$\mathchar''218$}}
     \raise 2.0pt\hbox{$\mathchar''13E$}}}
\newcommand{\beq}{\begin{equation}}
\newcommand{\eeq}{\end{equation}}

\lefthead{HAIMAN, MOHR \& HOLDER}
\righthead{COSMOLOGICAL PARAMETERS FROM GALAXY CLUSTER SURVEYS}

\begin{document}

\title{Constraints on Cosmological Parameters from Future \\
Galaxy Cluster Surveys}
\author{Zolt\'an Haiman\altaffilmark{1,2,3}, Joseph J. 
Mohr\altaffilmark{4,5,6} 
\& Gilbert P. Holder\altaffilmark{6}}
\altaffiltext{1}{Hubble Fellow}
\altaffiltext{2}{Princeton University Observatory, Princeton, NJ}
\altaffiltext{3}{NASA/Fermilab Astrophysics Center, Fermi National 
Accelerator Laboratory, Batavia, IL}
\altaffiltext{4}{Chandra Fellow}
\altaffiltext{5}{Departments of Astronomy and Physics, University of 
Illinois, 1202 W. Green St, Urbana, IL  61801}
\altaffiltext{6}{Department of Astronomy and Astrophysics, University of 
Chicago, 5640 S. Ellis Ave, Chicago, IL 60637}

\authoremail{zoltan@astro.princeton.edu}
\authoremail{jmohr@uiuc.edu}

\begin{abstract}
We study the expected redshift evolution of galaxy cluster abundance between
$0\lsim z \lsim 3$ in different cosmologies, including the effects of the
cosmic equation of state parameter $w\equiv p/\rho$.  Using the halo mass
function obtained in recent large scale numerical simulations, we model the
expected cluster yields in a 12~deg$^{2}$ Sunyaev-Zel'dovich Effect (SZE)
survey and a deep 10$^{4}$~deg$^{2}$ X-ray survey over a wide range of
cosmological parameters.  We quantify the statistical differences among
cosmologies using both the total number and redshift distribution of clusters.
Provided that the local cluster abundance is known to a few percent accuracy,
we find only mild degeneracies between $w$ and either $\Omega_m$ or $h$.  As a
result, both surveys will provide improved constraints on $\Omega_m$ and $w$.
The $\Omega_m$--$w$ degeneracy from both surveys is complementary to those
found either in studies of CMB anisotropies or of high--redshift Supernovae
(SNe).  As a result, combining these surveys together with either CMB or SNe
studies can reduce the statistical uncertainty on both $w$ and $\Omega_m$ to
levels below what could be obtained by combining only the latter two data sets.
Our results indicate a formal statistical uncertainty of $\approx 3\%$ (68$\%$
confidence) on both $\Omega_m$ and $w$ when the SZE survey is combined with
either the CMB or SN data; the large number of clusters in the X--ray survey
further suppresses the degeneracy between $w$ and both $\Omega_m$ and $h$.
Systematics and internal evolution of cluster structure at the present pose
uncertainties above these levels.  We briefly discuss and quantify the relevant
systematic errors.  By focusing on clusters with measured temperatures in the
X--ray survey, we reduce our sensitivity to systematics such as non-standard
evolution of internal cluster structure.

\end{abstract}

\section{Introduction}

It has long been realized that clusters of galaxies provide a uniquely useful
probe of the fundamental cosmological parameters.  The formation of the
large--scale dark matter (DM) potential wells of clusters is likely independent
of complex gas dynamical processes, star formation, and feedback, and involve
only gravitational physics.  As a result, the abundance of clusters $N_{\rm
tot}$ and their distribution in redshift $dN/dz$ should be determined purely by
the geometry of the universe and the power spectrum of initial density
fluctuations.  Exploiting this relation, the observed abundance of nearby
clusters has been used to constrain the amplitude $\sigma_{\rm 8}$ of the power
spectrum on cluster scales to an accuracy of $\sim 25\%$
(e.g. \cite{white93,viana96}).  The value of $\sigma_{\rm 8}$ in these studies
depends on the assumed underlying cosmology, especially on the density
parameters $\Omega_m$ and $\Omega_\Lambda$. Subsequent works
(\cite{bahcall98,blanchard98,viana99}) have shown that the redshift--evolution
of the observed cluster abundance places useful constrains on these two
cosmological parameters.

In the above studies, the equation of state for the $\Lambda$--component has
been implicitly assumed to be $p=w\rho$ with $w=-1$.  The recent suggestion
that $w$ might be different from $-1$, or even redshift dependent
(\cite{turner97,caldwell98}) has inspired several studies of cosmologies with a
component of dark energy.  From a particle physics point of view, such $w>-1$
can arise in a number of theories (see
\cite{freese87,ratra88,turner97,caldwell98} and references therein). It is
therefore of considerable interest to search for possible astrophysical
signatures of the equation of state, especially those that distinguish $w=-1$
from $w>-1$. Wang et al. (2000) has summarized current astrophysical
constraints that suggest $-1 \leq w \lsim -0.2$; while recent observations of
Type Ia SNe suggest the stronger constraint $w \lsim -0.6$ (\cite{ptw99}).

The galaxy cluster abundance provides a natural test of models that include a
dark energy component with $w\ne -1$, because $w$ directly affects the linear
growth of fluctuations $D_z$, as well as the cosmological volume element
$dV/dzd\Omega$. Furthermore, because of the dependence of the angular diameter
distance $d_A$ on $w$, the experimental detection limits for individual
clusters, e.g., from the Sunyaev--Zel'dovich effect (SZE) decrement or the
X--ray luminosity, depend on $w$.  Wang \& Steinhardt (1998, hereafter WS98)
studied the constraints on $w$ from a combination of measurements of the
cluster abundance and Cosmic Microwave Background (CMB) anisotropies.  Their
work has shown that the slope of the comoving abundance $dN/dz$ between $0<z<1$
depends sensitively on $w$, an effect that can break the degeneracies between
$w$ and combinations of other parameters $(h,\Omega,n)$ in the CMB anisotropy
alone.

Here we consider in greater detail the constraints on $w$, and other
cosmological parameters, from cluster abundance evolution.  Our main goals are:
(1) to quantify the statistical accuracy to which $w\neq -1$ models can be
distinguished from standard $\Lambda$ Cold Dark Matter (CDM) cosmologies using
cluster abundance evolution; (2) to assess these accuracies in two specific
cluster surveys: a deep SZE survey (\cite{carlstrom99}) and a large solid angle
X--ray survey, and (3) to contrast constraints from cluster abundance to those
from CMB anisotropy measurements and from luminosity distances to
high--redshift Supernovae (\cite{schmidt98,perlmutter99}).

Our work differs from the analysis of WS98 in several ways. We examine the
surface density of clusters $dN/dzd\Omega$, rather than the comoving number
density $n(z)$.  This is important from an observational point of view, because
the former, directly measurable quantity inevitably includes the additional
cosmology-dependence from the volume element $dV/dzd\Omega$.  We incorporate
the cosmology--dependent mass--limits expected from both types of surveys.
Because the SZE survey has a nearly $z$--independent sensitivity, we find that
high--redshift clusters at $z>1$ yield useful constraints, in addition to those
studied by WS98 in the range $0<z<1$. Finally, we quantify the statistical
significance of differences in the models by applying a combination of a
Kolmogorov--Smirnov (KS) and a Poisson test to $dN/dzd\Omega$, and obtain
constraints using a grid of models for a wide range of cosmological parameters.

This paper is organized as follows. In \S~\ref{sec:surveys}, we describe the
main features of the proposed SZE and X-ray surveys relevant to this work. In
\S~\ref{sec:models} we briefly summarize our modeling methods and assumptions.
In \S~\ref{sec:sensitivity}, we quantify the effect of individual variations of
$w$ and of other parameters on cluster abundance and evolution. In
\S~\ref{sec:wconstraints}, we obtain the constraints on these parameters by
considering a grid of different cosmological models. In
\S~\ref{sec:discussion}, we discuss our results and the implications of this
work.  Finally, in \S~\ref{sec:conclusions}, we summarize our conclusions.

\section{Cluster Surveys}
\label{sec:surveys}

The observational samples available for studies of cluster abundance evolution
will improve enormously over the coming decade.  The present samples of tens of
intermediate redshift clusters (e.g., \cite{gioia90,vikhlinin98}) will be
replaced by samples of thousands of intermediate redshift and hundreds of high
redshift ($z>1$) clusters.  At a minimum, the analysis of the European Space
Agency {\it X--ray Multi--mirror Mission (XMM)} archive for serendipitously
detected clusters will yield hundreds, and perhaps thousands of new clusters
with emission weighted mean temperature measurements (\cite{romer00}).
Dedicated X-ray and SZE surveys could likely surpass the {\it XMM} sample in
areal coverage, number of detected clusters or redshift depth.  The imminent
improvement of distant cluster data motivates us to estimate the cosmological
power of these future surveys.  Note that in practice, the only survey details
we utilize in our analyses are the virial mass of the least massive, detectable
cluster (as a function of redshift and cosmological parameters), and the solid
angle of the survey. We include here a brief description of two representative
surveys.

\subsection{A Sunyaev--Zel'dovich Effect Survey}
\label{subsec:SZ}

The SZE survey we consider is that proposed by Carlstrom and collaborators
(\cite{carlstrom99}).  This interferometric survey is particularly promising,
because it will detect clusters more massive than $\sim2\times10^{14}M_\odot$,
nearly independent of their redshift.  Combined, this low mass threshold and
its redshift independence produce a cluster sample which extends, depending on
cosmology, to redshifts $z\sim3$.  The proposed survey will cover 12~deg$^2$ in
a year; it will be carried out using ten 2.5~m telescopes and an 8~GHz
bandwidth digital correlator operating at cm wavelengths (\cite{mohr99}).  The
detection limit as a function of redshift and cosmology $M_{\rm
min}(z,\Omega_m,h)$ for this survey has been studied using mock observations of
simulated galaxy clusters (\cite{holder00}), and we draw on those results here.

Optical and near infrared followup observations will be required to
determine the redshifts of SZE clusters.  Given the relatively small
solid angle of the survey, it will be straightforward to obtain deep,
multiband imaging.  We expect that the spectroscopic followup will
require access to a multiobject spectrograph on a 10~m class
telescope.  The ongoing development of infrared spectrographs may
greatly enhance our ability to effectively measure redshifts for the
most distant clusters detected in the SZE survey.

\subsection{A Deep, Large Solid Angle X--ray Survey}
\label{subsec:COSMEX}

We also consider the cosmological sensitivity of a large solid angle, deep
X-ray imaging survey.  The characteristics of our survey are similar to those
of a proposed Small Explorer class mission, called the Cosmology Explorer,
spear-headed by G. Ricker and D. Lamb.  The survey depth is
$3.6\times10^6$~cm$^2$s at 1.5~keV, and the coverage is 10$^4$~deg$^2$
(approximately half the available unobscured sky).  We assume that the imaging
characteristics of the survey are sufficient to allow separation of the $10\%$
clusters from the $90\%$ AGNs and galactic stars.  We focus on clusters which
produce 500 detected source counts in the 0.5:6.0~keV band, sufficient to
reliably estimate the emission weighted mean temperature in a survey of this
depth (the external and internal backgrounds sum to $\sim1.4$~cts/arcmin$^2$).

To compute the number of photons detected from a cluster of a particular flux,
we assume the clusters emit Raymond--Smith spectra (\cite{raymond77}) with
$1\over3$ solar abundance, and we model the effects of Galactic absorption
using a constant column density of $n_H=4\times10^{20}$~cm$^{-2}$.  The
metallicity and Galactic absorption we've chosen are representative for a
cluster studied in regions of high Galactic latitude; when analyzing a real
cluster one would, of course, use the Galactic $n_{H}$ appropriate at the
location of the cluster.  Cluster metallicities vary, but for the 0.5:6~keV
band, line emission contributes very little flux for clusters with temperatures
above 2~keV.  For example, if the cluster metallicity were doubled to $2\over3$
solar, the conversion between flux and the observed counts in the
0.5:6~keV band for this particular survey would vary by $\sim1.4$\% and
$\sim0.1$\% for Raymond-Smith spectral models with temperatures $kT=2$~keV and
10~keV, respectively.  We assume that the detectors have a quantum efficiency
similar to the ACIS detectors (\cite{bautz98,chartas98}) on the {\it Chandra
X-ray Observatory}, and the energy dependence of the mirror effective area
mimics that of the mirror modules on ABRIXAS (\cite{friedrich98}).

The X-ray survey could be combined with the Sloan Digital Sky Survey
(SDSS) to obtain redshifts for the clusters -- the redshift
distribution of the clusters which produce 500 photons in the survey
described above is well sampled at the SDSS photometric redshift
limit.

\subsection{Determining the Survey Limiting Mass $M_{\rm min}$}
\label{subsec:mlim}

For our analysis, the most important aspect of both surveys is the
limiting halo mass $M_{\rm min} (z,\Omega_m,w,h)$, as a function of
redshift and cosmological parameters. More specifically, we seek the
relation between the detection limit of the survey, and the
corresponding limiting ``virial mass''.  In our modeling below, we
will be using the mass function of dark halos obtained in large scale
cosmological simulations (\cite{jenkins00}). In these simulations,
halos are identified as those regions whose mean spherical overdensity
exceeds the fixed value $\delta\rho/\rho_b=180$ (with respect to the
background density $\rho_b$, and irrespective of cosmology; see
discussion below).  In what follows, we adopt the same definition for
the mass of dark halos associated with galaxy clusters.

In the X-ray survey, $M_{\rm min}$ follows from the cluster X-ray
luminosity -- virial mass relation and the details of the survey.  We
adopt the relation between virial mass and temperature obtained in
hydrodynamical simulations by Bryan \& Norman (1998),
\begin{equation}
M_{\rm vir}=a\frac{T^{3/2}}{E(z)\sqrt{\Delta_c(z)}},
\label{eq:mt}
\end{equation}
where $H(z)=H_0 E(z)$ is the Hubble parameter at redshift $z$, $a=1.08$ is a
normalization determined from the hydrodynamical simulations, and $\Delta_c$ is
the enclosed overdensity (relative to the critical density) which defines the
cluster virial region.  The normalization $a$ is found to be relatively
insensitive to cosmological parameters, and the redshift evolution of
Equation~\ref{eq:mt} appears to be consistent with the hydrodynamical
simulations in those models where it has been tested (\cite{bryan98}).  Here we
assume that Equation~\ref{eq:mt} holds in all cosmologies with the same value
of $a$ (see $\S$\ref{subsec:systematics} for a discussion of the effects of
errors in the mass-temperature relation), and use the fitting formulae for
$\Delta_c$ provided by WS98, which includes the case $w\ne -1$.  Finally, we
convert $M_{\rm vir}$ from Equation~\ref{eq:mt} to the mass $M_{180}$ enclosed
within the spherical overdensity of $\delta\rho/\rho=180$ (with respect to the
background density), assuming that the halo profile is well described by the
NFW model with concentration $c=5$ (Navarro, Frenk \& White 1997, hereafter
NFW).

We next utilize Equation~\ref{eq:mt}, together with the relation between
bolometric luminosity and temperature found by Arnaud \& Evrard (1999), to find
the limiting mass of a cluster that produces 500 photons in the 0.5:6.0~keV
band in a survey exposure.  For these calculations we assume that the
luminosity-temperature relation does not evolve with redshift, consistent with
the currently available observations (\cite{mushotzky97}; relaxing this
assumption is discussed below in \S~\ref{sec:discussion}).

For an interferometric SZE survey, the relevant observable is the
cluster visibility $V$, which is the Fourier transform of the
cluster SZE brightness distribution on the sky as seen by the
interferometer.  The visibility is proportional to the total SZE flux
decrement $S_\nu$,
\begin{equation}
V\propto S_\nu (M,z)  \propto  f_{ICM}\frac{M\langle T_e\rangle_n}{d_A^2(z)} 
\label{eq:MlimSZ}
\end{equation}
where $\left<T_e\right>_n$ is the electron density weighted mean temperature,
$M$ is the virial mass, $f_{ICM}$ is the intracluster medium mass fraction and
$d_A$ is the angular diameter distance.  We normalize this relation using mock
observations of numerical cluster simulations (see \cite{mohr97} and
\cite{mohr99a}) carried out in three different cosmological models, including
noise characteristics appropriate to the proposed SZE array (see
\cite{holder00} for more details).  The ICM mass fraction is set to
$f_{ICM}=0.12$ in all three cosmological models.  This mass fraction is
consistent with analyses of X-ray emission from well defined samples if
$H_{0}=65$~km~s$^{-1}$~Mpc$^{-1}$, our fiducial value.  Note that we use the
same $f_{ICM}=0.12$ in all our cosmological models rather than varying it with
the $H_{0}$ scaling appropriate for analyses of cluster X-ray emission.  In the
discussion which follows, this choice allows us to focus solely on the
cosmological discriminatory power of cluster surveys; naturally, in
interpreting a real cluster survey one would likely allow $f_{ICM}$ to vary
with $H_{0}$.

\myputfigure{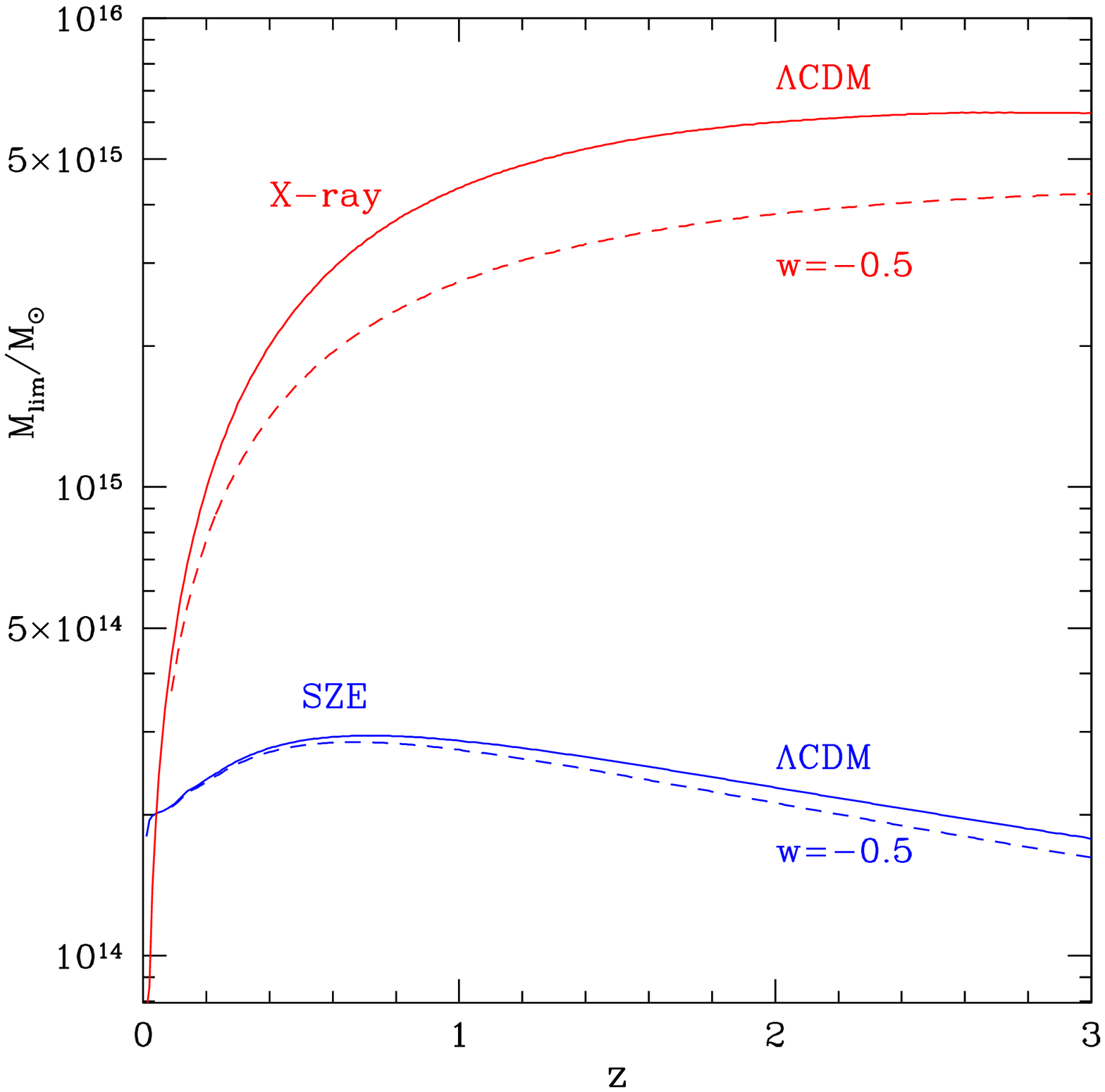}{3.2}{0.50}{-25}{-00}
\figcaption{Limiting cluster virial masses ($M_{180}$) for detection in the 
X--ray survey (upper pair of curves) and in the SZE survey (lower pair
of curves).  The solid curves show the mass limit in our fiducial flat
$\Lambda$CDM model, with $w=-1$, $\Omega_m=0.3$, and $h=0.65$, and the
dotted curves show the masses in the same model except with $w=-0.5$.
\label{fig:mlim}}
\vspace{\baselineskip} 

Note that for a flux limited survey, the limiting mass in
equation~\ref{eq:MlimSZ} is sensitive to cosmology through its
dependence on $d_A$ and the definition of the virial mass $M$.  We
adopt the simulation--normalized value of $M^{*}_{\rm min}(z)$ in our
fiducial cosmology as a template, and then we rescale this relation to
determine $M_{\rm min}(z)$ in the model of interest using the relation
\begin{equation}
M_{\rm min}(z)=
M^*_{min}(z)\frac{h^{*}}{h}\left[\frac{hd_A(z)}{h^{*}d^*_A(z)}\right]^{6/5}
\label{eq:Mscale}
\end{equation}
Here the superscript $^*$ refers to quantities in the $\Lambda$CDM
reference cosmology, and we have used the scaling of virial mass with
temperature (Eqn.~\ref{eq:mt}): $M\propto \left<T_e\right>_n^{3/2}$.
We tested this scaling by comparing it to mock observations in
simulations of two different cosmologies (open CDM and standard CDM),
and found that agreement was better than $\sim 10\%$ in the redshift
range $0<z<3$.  Finally, in the numerical simulations used to
calibrate Equation~\ref{eq:MlimSZ}, the halo mass was defined to be
the total mass enclosed within a region whose mean spherical interior
density is 200 times the critical density.  As in the X--ray case, we
convert $M_{\rm min}(z)$ from Equation~\ref{eq:Mscale} to the desired
mass $M_{180}$ by assuming that the halo profile follows NFW with 
concentration $c=5$.

The mass limits we derive for both surveys are shown in the redshift
range $0<z<3$ in Figure~\ref{fig:mlim}, both for $\Lambda$CDM and for
a $w=-0.5$ universe.  The SZE mass limit is nearly independent of
redshift, and changes little with cosmology. As a result, the cluster
sample can extend to $z\approx 3$. In comparison, the X--ray mass
limit is a stronger function of $w$, and it rises rapidly with
redshift. For the X-ray survey considered here the number of detected
clusters beyond $z\approx 1$ is negligible.

These mass limits incorporate some simplifying assumptions that have
not been tested in detail (although we consider small variations of
the mass limits below). Our goal is to capture the scaling with
cosmological parameters and redshift as best as presently possible.
However, we emphasize that further theoretical studies of the
sensitivities of these scalings to, for example, energy injection
during galaxy formation will be critical to interpreting the survey
data.  In the case of the X--ray survey, the cluster sample will have
measured temperatures, allowing the limiting mass to be estimated
independent of the cluster luminosity.  In the case of the SZE survey, deep X-ray
followup or multifrequency SZE followup observations should yield
direct measurements of the limiting mass.

\section{Estimating the Cluster Survey Yield}
\label{sec:models}

To derive cosmological constraints from the observed number
and redshift distribution of galaxy clusters, the fundamental quantity
we need to predict is the comoving cluster mass function.  The
Press--Schechter formalism (\cite{press74}; hereafter PS), which
directly predicts this quantity in any cosmology, has been shown to be
in reasonably good agreement (i.e. to within a factor of two)
with results of N--body simulations, in cosmologies and halo
mass ranges where it has been tested (\cite{lacey94,gross98,lee99}).
Numerical simulations have only recently reached the large size
required to accurately determine the mass function of the rarest, most
massive objects, such as galaxy clusters with $M>10^{15}M_{\odot}$.

In this paper, we adopt the halo mass function found in a series of recent
large--scale cosmological simulations by Jenkins et al.~2000.  The results of
these simulations are particularly well--suited for the present application.
The large simulated volumes allow a statistically accurate determination of the
halo mass function; for halo masses of interest here, to better than $\lsim
30\%$.  In addition, the mass function is computed in three different
cosmologies at a range of redshifts, and found to obey a simple 'universal'
fitting formula.  Although this does not guarantee that the same scaling holds
in other, untested cosmologies, we make this simplifying assumption in the
present paper.  In the future, the validity of this assumption has to be tested
by studying the numerical mass function across a wider range of cosmologies.

Generally, the simulation mass function predicts a significantly larger
abundance of massive clusters than does the PS formula.  For sake of
definiteness, we note that in the simulations, halos are identified as those
regions whose mean spherical overdensity exceeds the fixed value
$\delta\rho/\rho_b=180$ with respect to the background density $\rho_b$.  This
is somewhat different from the typical halo definition within the context of
the PS formalism, where the overdensity, relative to the critical density, is
taken to be that of a collapsing spherical top--hat at virialization.

Following Jenkins et al.~2000, we assume that the comoving number
density $(dn/dM)dM$ of clusters at redshift $z$ with mass $M\pm dM/2$
is given by the formula,
\beq
\frac{dn}{dM}(z,M)=
0.315 
\frac{\rho_0}{M}  
\frac{1}{\sigma_M}
\frac{d\sigma_M}{dM}
\exp\left[-\left|0.61-\log(D_z\sigma_M)\right|^{3.8}\right],
\label{eq:dndm}
\eeq
where $\sigma_M$ is the r.m.s. density fluctuation, computed on
mass--scale $M$ from the present--day linear power spectrum
(\cite{eisenstein98}), $D_z$ is the linear growth function, and
$\rho_0$ is the present--day mass density. The directly observable
quantity, i.e. the average number of clusters with mass above $M_{\rm
min}$ at redshift $z\pm dz/2$ observed in a solid angle $d\Omega$ is
then simply given by
\beq
\frac{dN}{dzd\Omega}\left(z\right) =
\left[
\frac{dV}{dzd\Omega}\left(z\right)
\int_{M_{\rm min}(z)}^\infty dM \frac{dn}{dM} 
\right]
\label{eq:dNdzdom}
\eeq 
where $dV/dzd\Omega$ is the cosmological volume element, and $M_{\rm
min}(z)$ is the limiting mass as discussed in section
\ref{subsec:mlim}.  Equations \ref{eq:dndm} and \ref{eq:dNdzdom}
depend on the cosmological parameters through $\rho_0$, $D_z$, and
$dV/dzd\Omega$, in addition to the mild dependence of $\sigma_M$ on
these parameters through the power spectrum (although the dependence
on the power--spectrum is more pronounced in the X--ray survey, where
the limiting mass varies strongly with redshift).  Note that the
comoving abundance $dn/dM$ is exponentially sensitive to the growth
function $D_z$. We use convenient expressions for $dV/dzd\Omega$ and
$D_z$ in open and flat $\Omega_\Lambda$ cosmologies available in the
literature (\cite{peebles80,carroll92,eisenstein96}).  In the case of
cosmologies with $w\neq -1$, we have evaluated $dV/dzd\Omega$
numerically, but used the fitting formulae for $D_z$ obtained by WS98,
which are accurate to better than 0.3\% for the cases of constant
$w$'s considered here.  

\subsection{Normalizing to Local Cluster Abundance}
\label{subsec:normalize}

To compute $dN/dzd\Omega$ from equation~\ref{eq:dNdzdom}, we must
choose a normalization for the density fluctuations $\sigma_M$.  This
is commonly expressed by $\sigma_8$; the present epoch, linearly
extrapolated {\it rms} variation in the density field filtered on
scales of $8h^{-1}$~Mpc.  To be consistent in our analysis, we choose
the normalization for each cosmology by fixing the local cluster
abundance above a given mass $M_{\rm nm}=10^{14} h^{-1}~M_\odot$.  In
all models considered, we set the local abundance to be
$1.03\times10^{-5}~(h/0.65)^3~{\rm Mpc^{-3}}$, the value derived in
our fiducial $\Lambda$CDM model (see below). We have chosen to
normalize using the local cluster abundance (upto a factor $h^3$)
above mass $M_{\rm nm}$ rather than above a particular emission
weighted mean temperature $kT_{\rm nm}$, because this removes the
somewhat uncertain cosmological sensitivity of the virial mass
temperature ($M-T_x$) relation from the normalization process;
spherical tophat calculations suggest a significant offset in the
$M-T_x$ normalization of the open and flat $\Omega_m=0.3$ models which
hydrodynamical simulations do not seem to reproduce
(\cite{evrard96,bryan98,viana99}).

An alternative approach to the above is to regard $\sigma_8$ a
``free--parameter'', on equal footing with the other parameters we let
float below. This possibility will be discussed further in \S~6.  Here
we note that our normalization approach is sensible, because the
number density of nearby clusters can be measured to within a factor
of $h^3$, and the masses of nearby clusters can be measured directly
through several independent means; these include the assumption of
hydrostatic equilibrium and using X-ray images and intracluster medium
(ICM) temperature profiles, weak lensing, or galaxy dynamical mass
estimates.  The only cosmological sensitivity of these mass estimators
is their dependence on the Hubble parameter $h$; we include this $h$
dependence when normalizing our cosmological models.  Note that
previous derivations of $\sigma_8$ (e.g. Viana \& Liddle 1993; Pen
1998) in various cosmologies from the local cluster abundance $N(>kT)$
above a fixed threshold temperature $kT_{\rm min}\sim 7$keV yielded a
constraint with the approximate scaling $\sigma_8\Omega_m^{1/2}
\approx 0.5$.  We find a similar relation when varying $\Omega_m$
away from our fiducial cosmology; however, we note that if a $\sim 5$
times smaller threshold temperature were used, the constrained
combination would be quite different, $\sigma_8\Omega_m\sim$constant.
Since our adopted normalization is based on mass, rather than
temperature, in general, we find still different scalings. As an
example, when $h=0.65$ and $w=-1$ are kept fixed, our normalization
procedure translates into $\sigma_8(\Omega_m/0.3)^{0.85}\approx 0.9$.

\subsection{Fiducial Cosmological Model}

The parameters we choose for of our fiducial cosmological model are
$(\Omega_\Lambda,\Omega_{\rm m},h,\sigma_{8},n)=(0.7,0.3,0.65,0.9,1)$.
This flat $\Lambda$CDM model is motivated as a ``best--fit'' model
that produces a local cluster abundance consistent with observations
(\cite{viana99}), and satisfies the current constraints from CMB
anisotropy (\cite{lange00}, see also \cite{white00}), high--$z$ SNe, and
other observations (Bahcall et al. 1999).  We have assumed a baryon
density of $\Omega_b h^2=0.02$, consistent with recent D/H
measurements (e.g. Burles \& Tytler 1998).  Note that the power
spectrum index $n$ is not important for the analysis presented here,
because we normalize on cluster scales $\sigma_{8}$, and we find that
this minimizes the effect of varying $n$ on the density fluctuations
relevant to cluster formation.

\section{Exploring the Cosmological Sensitivity}
\label{sec:sensitivity}

In this section, we describe how variations of the individual
parameters $\Omega$, $w$, and $h$, as well as the cosmological
dependence of the limiting mass $M_{\rm min}$, affect the cluster
abundance and redshift distribution.  This will be useful in
understanding the results of the next section, when a full grid of
different cosmologies is considered.  We then describe our method of
quantifying the statistical significance of differences between the
distributions $dN/dz$ in a pair of different cosmologies.

\subsection{Single Parameter Variations}
\label{subsec:parameters}

The surface density of clusters more massive than $M_{\rm min}$
depends on the assumed cosmology mainly through the growth function
$D(z)$ and volume element $dV/dzd\Omega$, as well as through the
cosmology--dependence of the limiting mass $M_{\rm min}$ itself.  In
the approach described in section \ref{sec:models}, once a cosmology
is specified, the normalization of the power spectrum $\sigma_{8}$ is
found by keeping the abundance of clusters at $z=0$ constant.  We
therefore consider only three ``free'' parameters, $w$, $h$,
$\Omega_m$, specifying the cosmology.  We assume the universe to be
either flat ($\Omega_Q=1-\Omega_m$), or open with $\Omega_Q=0$.

\myputfigure{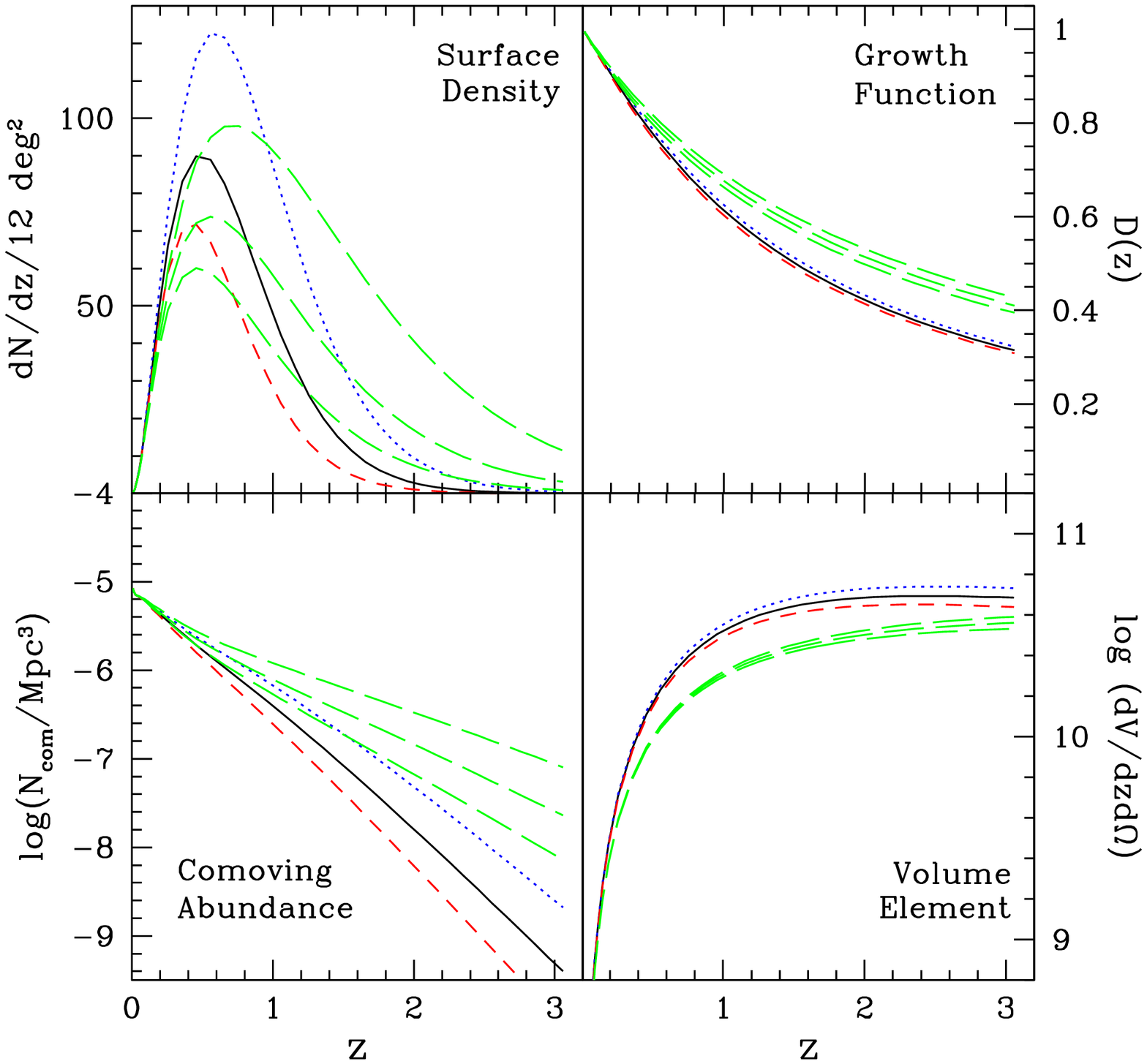}{3.2}{0.49}{-25}{-00} 
\figcaption{Effect of changing $\Omega_m$ when all other parameters are held
fixed.  The four panels show (clockwise from upper left) the surface
density of clusters at redshift $z$; the linear growth function; the
volume element in units of ${\rm Mpc^3~sr^{-1}~redshift^{-1}}$; and
the comoving cluster abundance.  The solid curve shows our fiducial
flat $\Lambda$CDM model, with $w=-1$, $\Omega_m=0.3$, and $h=0.65$.
Also shown are models with $\Omega=0.27$ (dotted curve); $\Omega=0.33$
(short--dashed curve); and OCDM models with $\Omega =0.27,0.30,0.33$
(long--dashed curves, top to bottom).
\label{fig:clust_om}}

\myputfigure{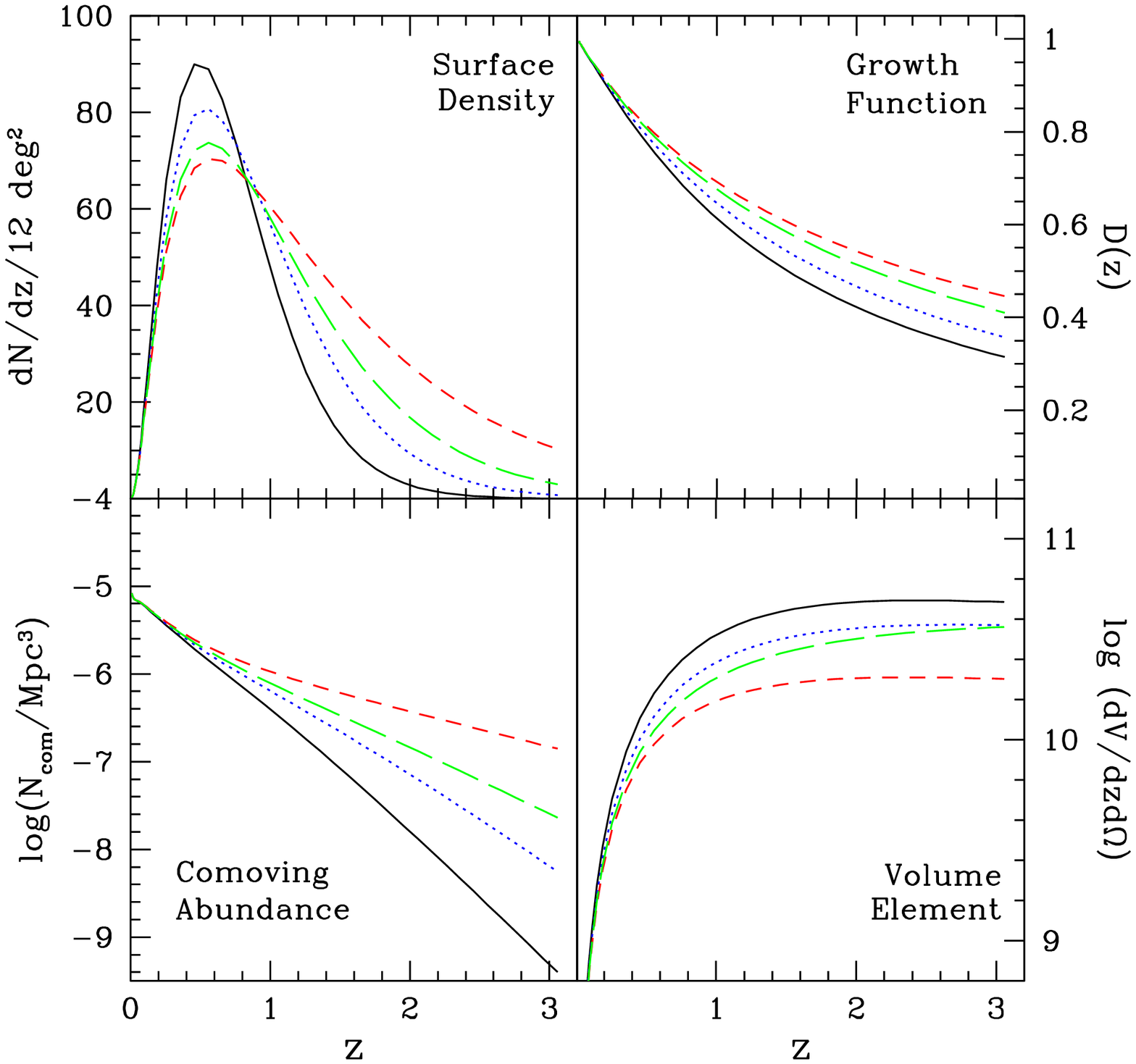}{3.2}{0.49}{-25}{-00}
\figcaption{Effect of changing $w$ when all other parameters are held 
fixed.  The solid curve shows our fiducial flat $\Lambda$CDM model,
with $w=-1$, $\Omega_m=0.3$, and $h=0.65$. The dotted curve is the
same model with $w=-0.6$, the short--dashed curve with $w=-0.2$, and
the long--dashed curve is an open CDM model with $\Omega_m=0.3$.
\label{fig:clust_w}}

\subsubsection{Changing $\Omega_m$}

The effects of changing $\Omega_m$ are demonstrated in
Figure~\ref{fig:clust_om}.  The curves correspond to a flat $\Lambda$CDM
universe with ($h=0.65,w=-1$), and $\Omega_m=0.27$ (dotted), $\Omega_m=0.30$
(solid), and $\Omega_m=0.33$ (short--dashed).  In addition, the long--dashed
curves show the same three models (top to bottom), assuming open CDM with
$\Omega_\Lambda=0$.  The top left panel shows the total number of clusters in a
12 square degree field, detectable down to the constant SZE decrement $S_{\rm
min}$.  As discussed in section \ref{subsec:mlim} above, a constant $S_{\rm
min}$ implies a redshift and cosmology--dependent limiting mass $M_{\rm min}$.
In the SZE case, we find that if we had not included this effect, the
sensitivity to $\Omega_m$ would have been somewhat stronger.  Several
conclusions can be drawn from Figure~\ref{fig:clust_om}. Overall, the top left
panel shows that a decrease in $\Omega_m$ increases the number of clusters (and
vice versa) at all redshifts.  Note that the dependence is strong, for
instance, a $10\%$ decrease in $\Omega_m$ increases the total number of
clusters by $\sim30\%$ in either $\Lambda$CDM or OCDM cosmologies.  As
emphasized by Bahcall \& Fan (1998), Viana \& Liddle (1999) and others, this
makes it possible to estimate an upper limit on $\Omega_m$ using current,
sparse data on cluster abundances (i.e. only a few high--$z$ clusters).  A
second important feature seen in the top left panel is that the shape of the
redshift distribution is not changed significantly, a conclusion that holds
both in $\Lambda$CDM and OCDM.  Finally, the remaining three panels reveal that
the effects of $\Omega_m$ arise mainly from the changes in the comoving
abundance (bottom left panel).  In flat $\Lambda$CDM, $\Omega_m$ has relatively
little effect on the volume or the growth function, and the comoving abundance
is determined by the value of $\sigma_8$ that keeps the the local abundance
constant at $z=0$ (we find $\sigma_8$=0.83 for $\Omega_m=0.33$ and
$\sigma_8=1.00$ for $\Omega_m=0.27$). In addition, we find that the change in
the shape of the underlying power spectrum with $\Omega_m$ enhances the
differences caused by $\Omega_m$ (when we artificially keep the power spectrum
at its $\Omega_m=0.3$ shape, we find $\sigma_8$=0.84 for $\Omega_m=0.33$).  We
also note that the volume element and the comoving abundance act in the same
direction: a lower $\Omega_m$ increases both the comoving abundance and the
volume element.  In OCDM, the growth function has a larger effect, and relative
to $\Lambda$CDM, the redshift distribution is much flatter.

\myputfigure{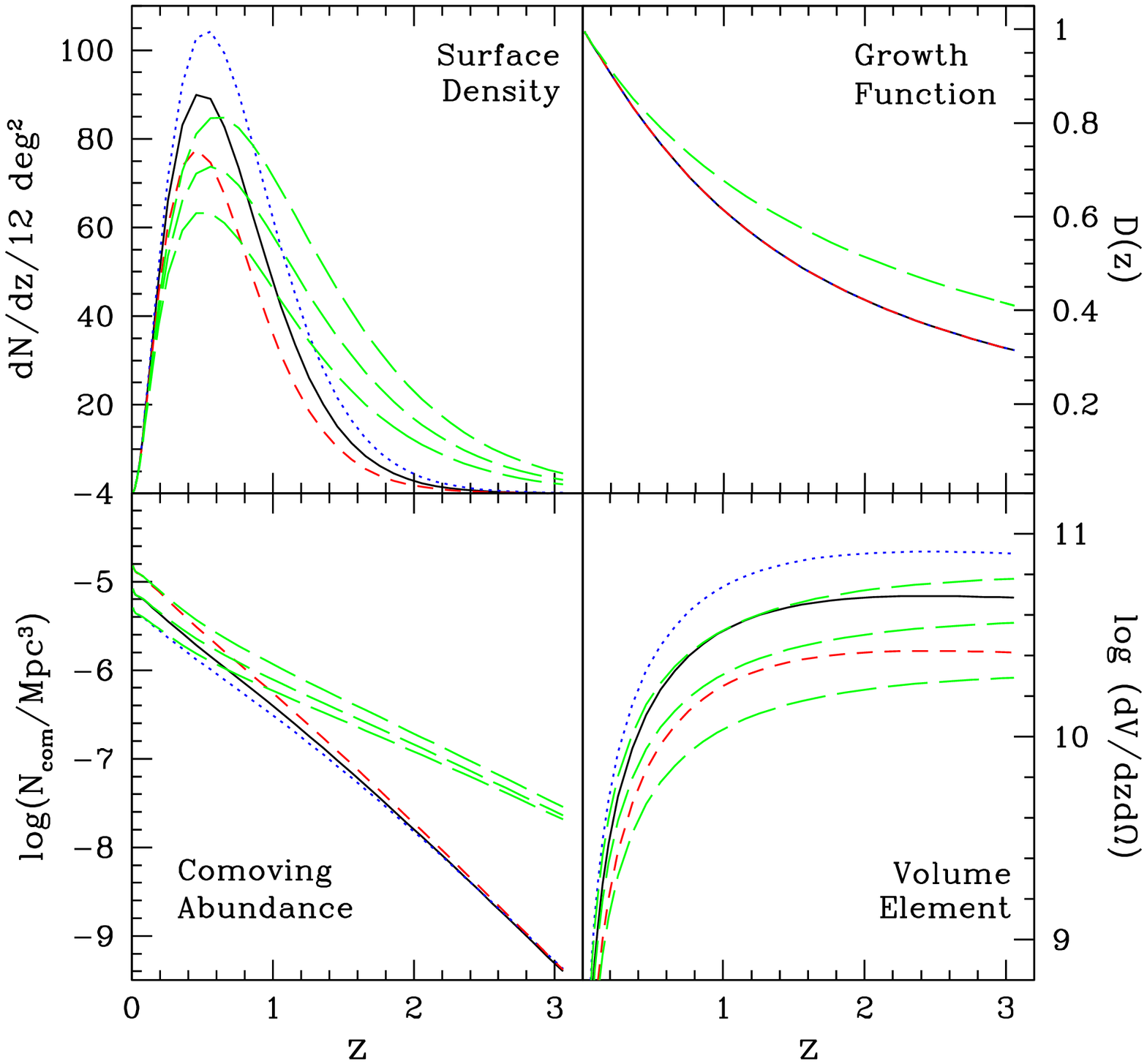}{3.2}{0.49}{-25}{-00}
\figcaption{Effect of changing $h$ when all other parameters are held fixed.
The $\Lambda$CDM model of Figure~\ref{fig:clust_w} is shown (solid
curve) together with models with $h=0.55$ (dotted curve); $h=0.80$
(short--dashed curve); and OCDM models with $h=0.55,0.65,0.80$
(long--dashed curves, top to bottom).
\label{fig:clust_h}}

\subsubsection{Changing $w$} 

The effects of changing $w$ are demonstrated in Figure~\ref{fig:clust_w}.  The
figure shows models with ($\Omega_m=0.3,h=0.65$) and with three different
$w$'s: $w=-1$ (solid curve), $w=-0.6$ (dotted curve), and $w=-0.2$
(short--dashed curve).  In addition, we show the result from an open CDM model
with ($\Omega=0.3,h=0.65$; long--dashed curve).  The figure reveals that
increasing $w$ above $w=-1$ causes the slope of the redshift distribution above
$z\approx 0.5$ to flatten, increasing the number of high--$z$ clusters.
Furthermore, ``opening'' the universe has an effect similar to increasing $w$.
The other three panels demonstrate the reason for these scalings.  The top
right panel shows that the growth function is flatter in higher $w$ models,
significantly increasing the comoving number density of high--redshift clusters
(bottom left panel).  The volume element (bottom right panel) has the opposite
behavior, in the sense the volume in higher--$w$ models is smaller, which tends
to balance the increase in the comoving abundance caused by the growth function
in the range $0<z\lsim 0.5$; but for higher redshifts, the growth function
``wins''. An important conclusion seen from Figure~\ref{fig:clust_w} is that
both the total number of clusters as well as the shape of their redshift
distribution, significantly depends on $w$.  We also note that in the SZE case,
our sensitivity to $w$ has been enhanced by the cosmological dependence of the
mass limit (opposite to what we found for the $\Omega_m$--sensitivity, which we
found was weakened by the same effect).

\subsubsection{Changing $h$}

Figure~\ref{fig:clust_h} demonstrates the effects of changing $h$.  Three
$\Lambda$CDM models are shown with ($\Omega_m=0.30,w=-1$), and $h=0.55$ (dotted
curve), $h=0.65$ (solid curve), and $h=0.80$ (short--dashed curves).  The
long--dashed curves correspond to OCDM models with the same parameters (top to
bottom).  Comparing the top right panel with that of Figure~\ref{fig:clust_om},
the qualitative behavior of $dN/dz$ under changes in $h$ and $\Omega_m$ are
similar: decreasing $h$ increases the total number of clusters, but does not
considerably change their redshift distribution.  However, the sensitivity to
$h$ is significantly less: the total number of clusters is seen to increase by
$\sim 25\%$ only when $h$ is decreased by the same percentage.  Note that the
growth function is not effected by $h$, and the $h$ sensitivity is driven by
our normalization process, which fixes the abundance at $z=0$
(see~\S~\ref{subsec:normalize}).  Since the volume scales as $\propto h^{-3}$,
we fix the comoving abundance to be proportional to $\propto h^{3}$.  As a
result, $dN/dzd\Omega$ is nearly independent of $h$.  In fact, the entire
$h$--dependence is attributable to the small change caused by $h$ in the shape
of the power spectrum (for a pure power--law spectrum, there would be no
$h$--dependence, and the three curves for the flat universe in the top left
panel of Figure~\ref{fig:clust_h} would look identical).

\myputfigure{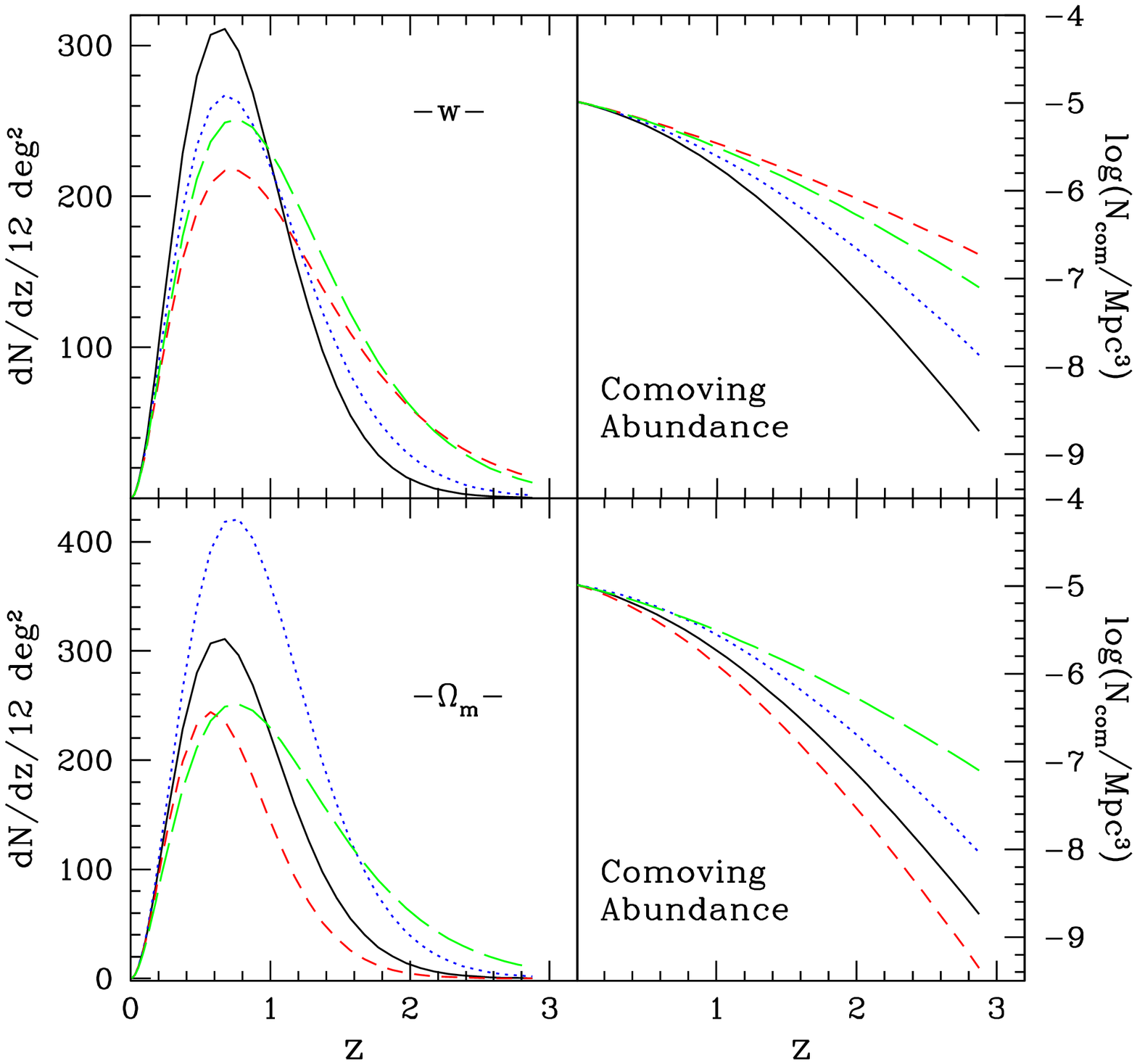}{3.2}{0.49}{-25}{-00} 
\figcaption{Effect of changing $w$ (upper panels) or $\Omega_m$ (lower panels)
when all other parameters are held fixed, including the mass limit.  The types
of the curves correspond to the different models in the SZE survey, as shown in
Figures~\ref{fig:clust_om}~\&~\ref{fig:clust_w}.
\label{fig:clust_mlim}}

\subsubsection{Abundances in the X--ray Survey}

The evolution of the cluster abundance, and its sensitivity to $\Omega_m$ and
$w$ in the X--ray survey are shown in Figure~\ref{fig:clust_x}.  Because of the
much larger solid angle surveyed, the numbers of clusters is significantly
larger than in the SZE case, despite the higher limiting mass
(cf. Fig~\ref{fig:mlim}).  Nevertheless, the general trends that can be
identified in the X--ray sample are similar to those in the SZE case. Raising
$w$ increases the total number of clusters, and flattens their redshift
distribution.  As in the SZE survey, raising $\Omega_m$ decreases the total
number of clusters.

\subsection{Effects of the Limiting Mass Function}

Finally, we examine the extent to which the above conclusions depend on the
cosmology and redshift--dependence of the limiting mass $M_{\rm min}$.  

\myputfigure{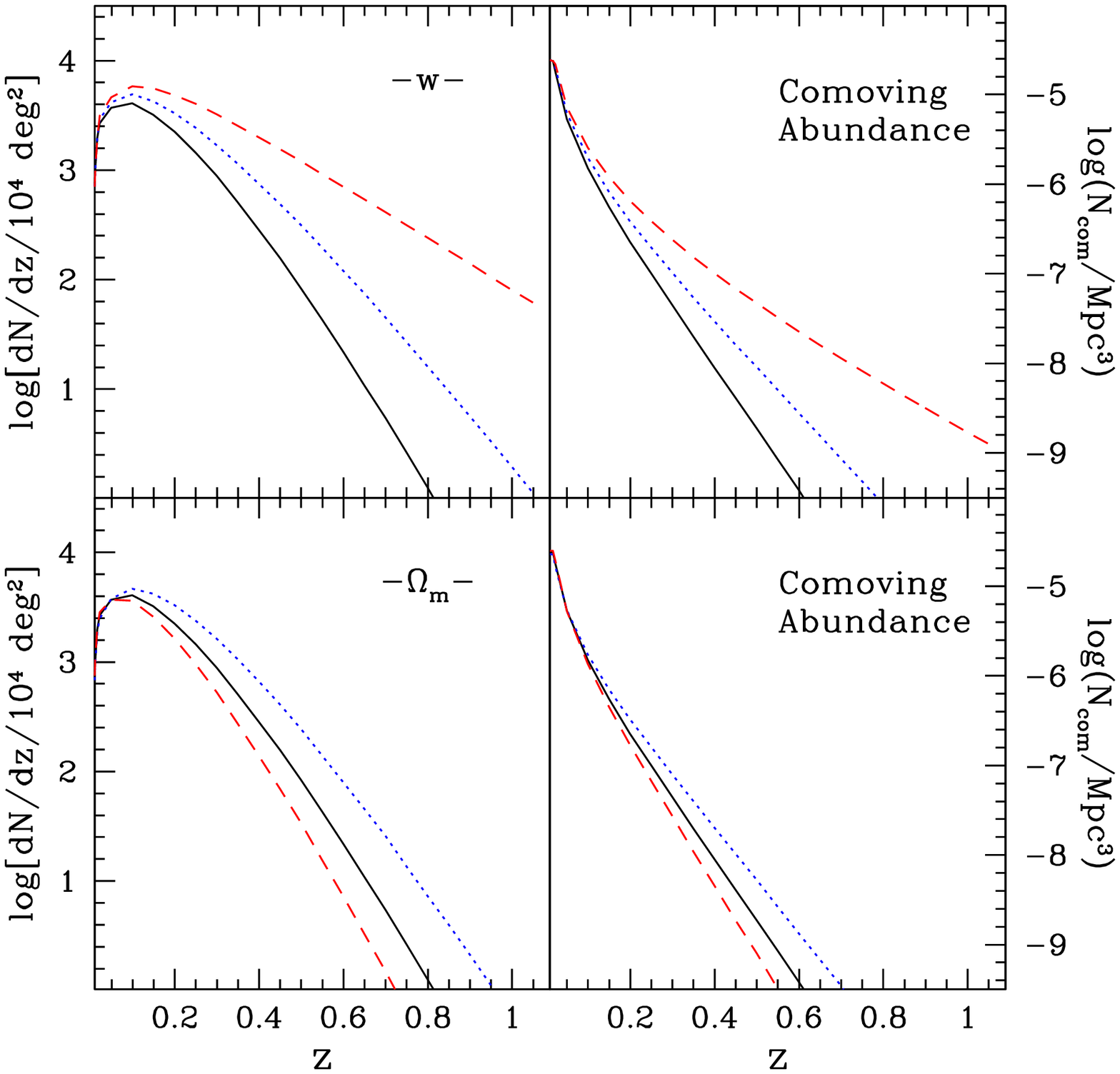}{3.1}{0.51}{-25}{-00} 
\figcaption{Effect of changing $w$ (upper panels) or $\Omega_m$ (lower panels)
when all other parameters are held fixed in the X--ray survey.  Note the much
larger numbers of clusters in comparison to the SZE survey. In the top panel,
the curves correspond to $w=-1$ (solid), $w=-0.6$ (dotted) and $w=-0.2$
(dashed).  In the bottom panel, the curves correspond to $\Omega_m=0.3$
(solid), $\Omega_m=0.27$ (dotted) and $\Omega_m=0.33$ (dashed).
\vspace{\baselineskip}
\label{fig:clust_x}}

\subsubsection{The SZE Survey}

We first compute cluster abundances above the fixed mass $M_{\rm
min}=10^{14}h^{-1}{\rm M_\odot}$, characteristic of the SZE survey detection
threshold in the range of cosmologies and redshifts considered here. The
results are shown in Figure~\ref{fig:clust_mlim}: the bottom panels show the
surface density and comoving abundance when $\Omega_m$ is changed (the models
are the same as in Figure~\ref{fig:clust_om}), and the top panels show the same
quantities under changes in $w$ (the cosmological models are the same as in
Figure~\ref{fig:clust_w}).  A comparison between Figures~\ref{fig:clust_mlim}
and \ref{fig:clust_w} gives an idea of the importance of the mass limit. The
general trend seen in Figure~\ref{fig:clust_w} remains true, i.e. increasing
$w$ flattens the redshift distribution at high--$z$.  However, when a constant
$M_{\rm min}$ is assumed, the ``pivot point'' moves to slightly higher
redshift, and the total number of clusters becomes less sensitive to $w$.
Similar conclusions can be drawn from a comparison of Figure~\ref{fig:clust_om}
with the bottom two panels of Figure~\ref{fig:clust_mlim}: under changes in
$\Omega_m$ the general trends are once again similar, but the differences
between the different models are amplified when a constant $M_{\rm min}$ is
used.  In summary, we conclude that in the SZE case (1) the variation of the
mass limit with redshift and cosmology has a secondary importance, and (2) it
weakens the $\Omega_m$ dependence, but strengthens the $w$ dependence.

\subsubsection{The X--ray Survey}

In comparison to the SZE survey, the X--ray mass limit is not only higher, but
is also significantly more dependent on cosmology (cf. Fig~\ref{fig:mlim}).  On
the other hand, the X--ray sample goes out only to the relatively low redshift
$z=1$, where the growth functions in the different cosmologies diverge
relatively little.  This suggests that in the X--ray case the mass limit is
more important than in the SZE survey. In order to separate the effects of the
changing mass limit from the change in the growth function and the volume
element, in Figure~\ref{fig:clust_xnoM} we show the sensitivity of $dN/dz$ to
changes in $\Omega_m$ and $w$, without including the effects from the mass
limit.  The same models are shown as in Figure~\ref{fig:clust_x}, except we
have artificially kept the mass limit at its value in the fiducial cosmology.
The figure reveals that essentially all of the $w$--sensitivity seen in
Figure~\ref{fig:clust_x} is caused by the changing mass limit; when $M_{\rm
min}$ is kept fixed, the cluster abundances change very little.  On the other
hand, comparing the bottom panels of Figures~\ref{fig:clust_x}
and~\ref{fig:clust_xnoM} shows that including the scaling of the mass limit
somewhat reduces the $\Omega_m$ dependence, just as in the SZE case.

\myputfigure{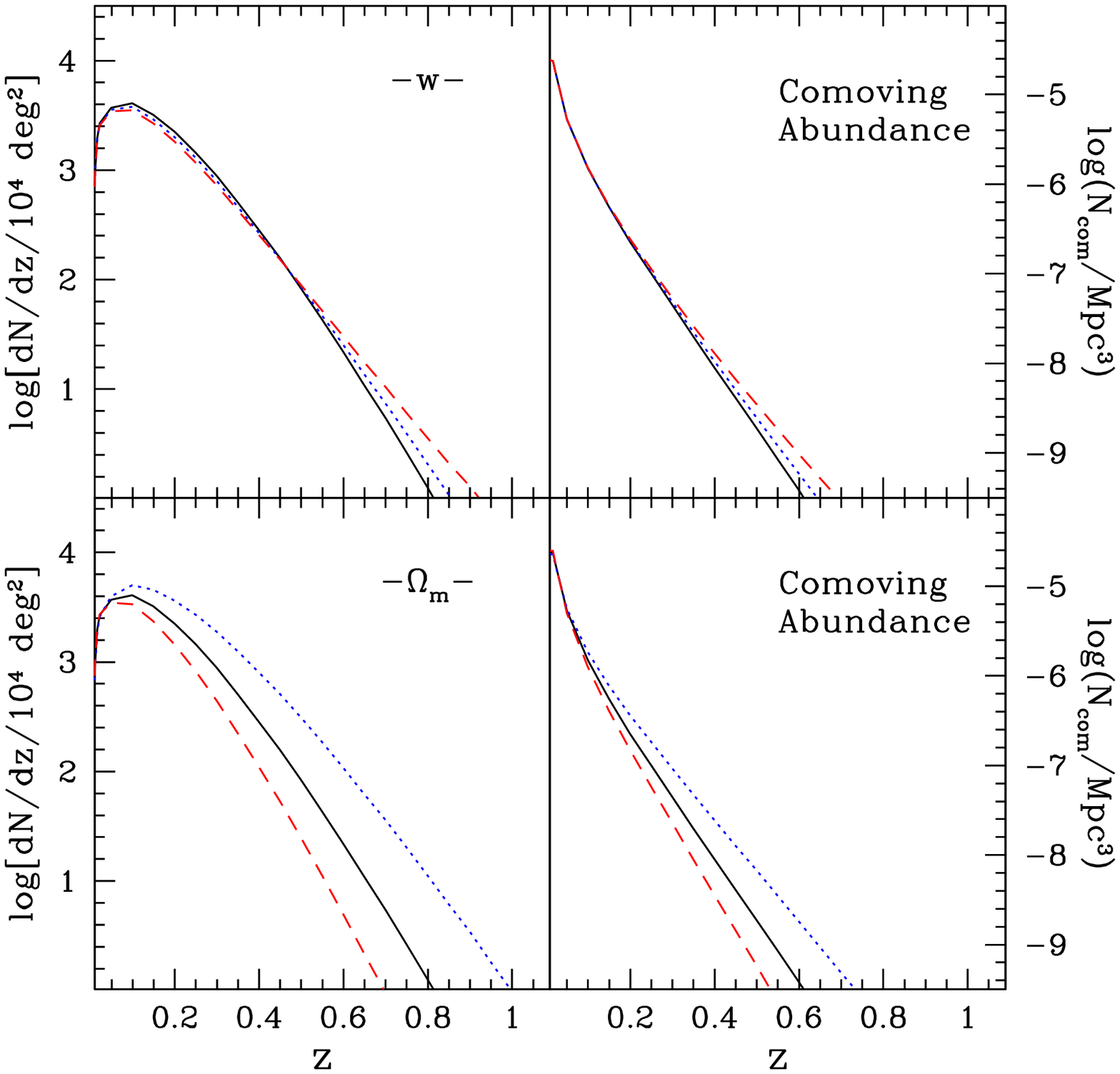}{3.1}{0.51}{-25}{-00} 
\figcaption{Effect of changing $w$ (upper panels) or $\Omega_m$ (lower panels)
when all other parameters are held fixed in an X--ray survey, and the survey
mass limit is held fixed at its fiducial value, irrespective of cosmology.  A
comparison with Figure~\ref{fig:clust_x} shows that nearly all of the
$w$--sensitivity is accounted for by the cosmology--dependence of the limiting
mass. On the other hand, the $\Omega_m$--sensitivity is caused mostly by the
growth function.
\vspace{\baselineskip}
\label{fig:clust_xnoM}}

\begin{figure*}[htb]
\vskip-0.2in
\hbox to \hsize{\hfil\hskip-0.2in\vbox to 3.3in{
\epsfysize=3.3in
\epsfbox{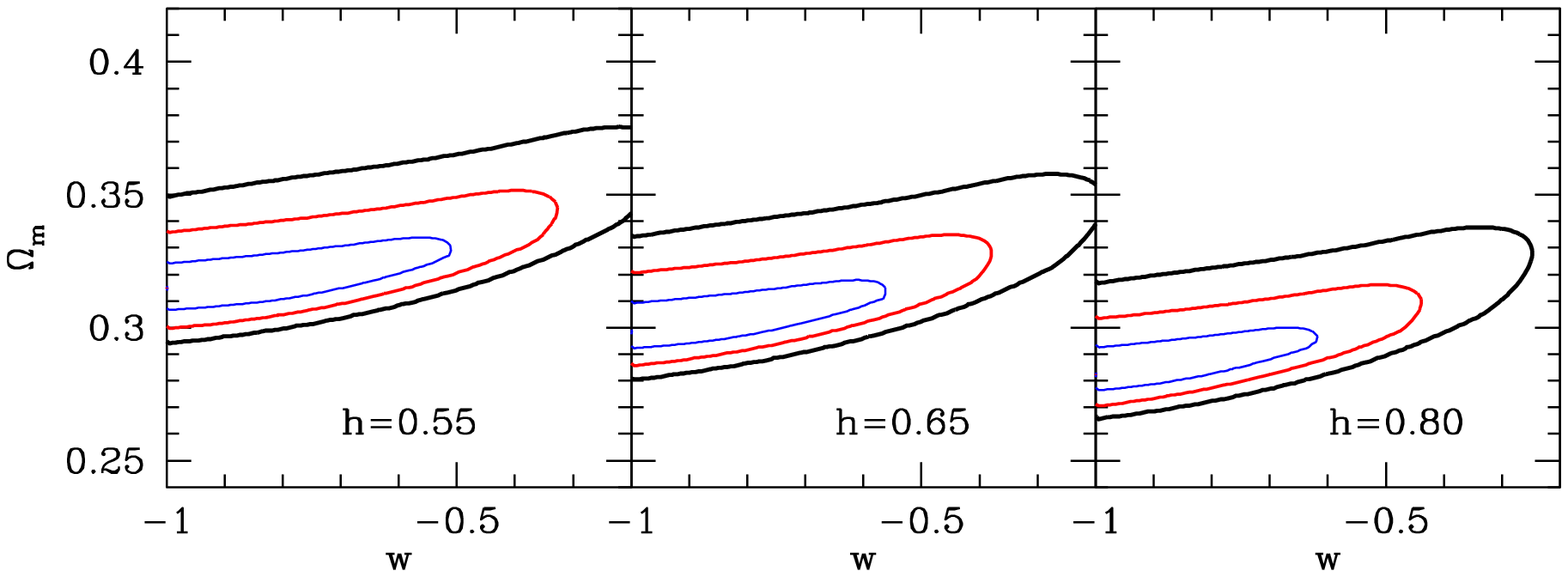}\vfil}\hfil}\vskip-0.4in
\caption{\footnotesize Contours of 1, 2, and 3$\sigma$ likelihood for different
models when they are compared to a fiducial flat $\Lambda$CDM model with
$\Omega_m=0.3$ and $h=0.65$, using the SZE survey.  The three panels show three
different cross--sections of constant total probability at fixed values of $h$
(0.55,0.65, and 0.80) in the investigated 3--dimensional $\Omega_m,w,h$
parameter space.}
\label{fig:slices}
\vskip-0.2in
\end{figure*}

\subsection{Overview of Cosmological Sensitivity}

In summary, we conclude that changes in $w$ modify both the normalization and
the shape of the redshift distribution of clusters, while changes in $\Omega_m$
or $h$ effect essentially only the overall amplitude.  This suggests that
changes in $w$ can not be fully degenerate with changes in either $\Omega_m$ or
$h$ (or a combination), making it possible to measure $w$ from cluster
abundances alone.  These conclusions hold either for clusters above a fixed
detection threshold in and SZE or X-ray survey, or for a sample of clusters
above a fixed mass.  We find that the sensitivity to $\Omega_m$ arises mostly
through the growth function, both in the SZE and X--ray surveys. This
sensitivity is slightly weakened by the scaling of the limiting mass $M_{\rm
min}$ with $\Omega_m$.  We find that the $w$ sensitivity is also dominated by
the growth function in the SZE survey, which goes out to relatively high
redshifts; but the sensitivity to $w$ is enhanced by the $w$--dependence of
$M_{\rm min}$.  In comparison, in the X--ray survey, which only probes
relatively low redshifts, nearly all of the $w$--sensitivity is caused by the
cosmology--dependence of the limiting mass, rather than the growth function.

\begin{figure*}[b]
\vskip-0.3in
\hbox to \hsize{\hfil\hskip-0.30in\vbox to 3.3in{
\epsfysize=3.3in
\epsfbox{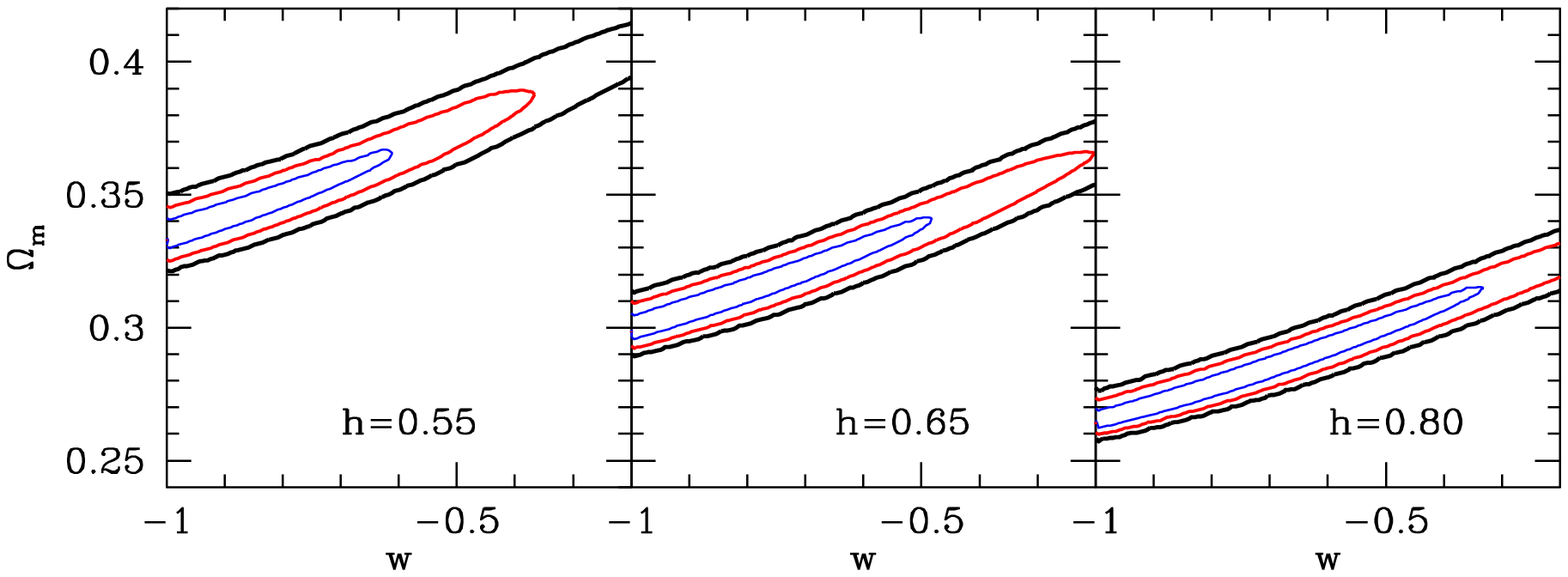}\vfil}\hfil}\vskip-0.40in
\caption{\footnotesize Contours of 1, 2, and 3$\sigma$ likelihood for models
when they are compared to a fiducial flat $\Lambda$CDM model, as in
Figure~\ref{fig:slices}, but for the X--ray survey.}
\label{fig:slicex}
\end{figure*}

\section{Constraints on Cosmological Parameters}
\label{sec:wconstraints}

We derive cosmological constraints by considering a 3--dimensional
grid of models in $\Omega_m,h$, and $w$.  As described above, we first
find $\sigma_8$ in each model, so that all models are normalized to
produce the same local cluster abundance at $z=0$.  We then compute
$dN/dzd\Omega$ in these models for $0.2 \leq \Omega_m \leq 0.5$, $0.5
\leq h \leq 0.9$, and $-1 \leq w \leq -0.2$.  The range for $w$
corresponds to that allowed by current astrophysical observations
(\cite{wang00}); although recent observations of Type Ia SNe suggest
the stronger constraint $w \lsim -0.6$ (\cite{ptw99}).

\subsection{Comparing $dN/dz$ in Two Different Cosmologies}
\label{subsec:probabilities}

The main goal of this paper is to quantify the accuracy to which $w$ can be
measured in future SZE and X--ray surveys.  To do this, we must answer the
following question: given a hypothetical sample of $N_{\rm tot}$ clusters (with
measured redshifts) obeying the distribution $dN_A/dz$ of the test model (A)
cosmology, what is the probability $P_{\rm tot}(A,B)$ that the same sample of
clusters is detected in the fiducial (B) cosmology, with distribution
$dN_B/dz$?  We have seen in section \ref{subsec:parameters} that the overall
amplitude, and the shape of $dN/dz$ are both important.  Motivated by this, we
define
\beq
P_{\rm tot}(A,B) = P_{0}(A,B) \times P_z(A,B)
\label{eq:probs}
\eeq 
where $P_{0}(A,B)$ is the probability of detecting $N_{A, \rm tot}$
clusters when the mean number is $N_{B, \rm tot}$, and $P_z(A,B)$ is
the probability of measuring the redshift distribution of model (A) if
the true parent distribution is that of model (B).  We assume $P_{0}$
is given by the Poisson distribution, and we use the
Kolmogorov--Smirnov (KS) test to compute $P_z(A,B)$ (\cite{press92}).
The main advantage of this approach, when compared to the usual
$\chi^2$ tests, is that we do not need to bin the data in redshift.

For reference, it is useful to quote here some examples for the
probabilities, taking ($\Omega_m=0.3,h=0.65,w=-1$) as the fiducial (B)
model.  For example, closest to this model in Figure \ref{fig:clust_w}
is the one with $w=-0.6$.  For this case, we find $P_0=0.25$ and
$P_z=0.1$ for a total probability of $P_{\rm tot}=0.025$.  In other
words, the two cosmologies could be distinguished at a likelihood of
$1.2\sigma$ using only the total number of clusters, at $1.6\sigma$
using only the shape of the redshift distribution, and at the
$2.3\sigma$ level using both pieces of information.  In this case, the
distinction is made primarily by the different redshift distributions,
rather than the total number of detected clusters.  Taking the
$\Omega_m=0.33$ $\Lambda$CDM cosmology from Figure \ref{fig:clust_om}
as another example for model (A), we find $P_0=0.0075$ (=$2.7\sigma$),
$P_z=0.78$ (=$0.3\sigma$), and a total probability of $P_{\rm
tot}=0.0058$ (=$2.8\sigma$). Not surprisingly, the shape of the
redshift distribution does not add significantly to the statistical
difference between these two models, which differ primarily by the
total number of clusters.

\subsection{Expectations from the Sunyaev--Zel'dovich Survey}
\label{subsec:wSZ}

Figure~\ref{fig:slices} shows contours of 1, 2, and 3$\sigma$ for the
total probability $P_{\rm tot}$ for models when compared to the
fiducial flat $\Lambda$CDM model.  For reference, we note that the
total number of clusters in the SZE survey in our fiducial model is
$\approx 100$, located between $0<z<3$. The three panels show three
different cross--sections of the investigated 3--dimensional
$\Omega_m,h,w$ parameter space, taken at constant values of $h=$ 0.55,
0.65, and 0.80, spanning the range of values preferred by other
observations.  The most striking feature in this figure is the
direction of the contours, which turn upwards in the $w,\Omega_m$
plane, and become narrower for larger values of $w$.  We find that the
trough of maximum probability for fixed $h=0.65$ is well described by
\beq 
(\Omega_m - 0.3) (w+1)^{-5/2} = 0.1,
\label{eq:degen}
\eeq with further constant shifts in $\Omega_m$ caused by changing $h$. The
$\pm 3\sigma$ width enclosed by the contours around this relation is relatively
narrow in $\Omega_m$ ($\pm 10\%$).  In a $\Lambda$CDM case, even when a large
range of values is considered for $h$ ($0.45 < h < 0.90$), the constraint $0.26
\lsim \Omega_m \lsim 0.36$ follows; when $w\ne -1$ is considered, the allowed
range widens to $0.27 \lsim \Omega_m \lsim 0.41$.  On the other hand, a wide
range of $w$'s is seen to be consistent with $w=-1$: the largest value shown,
$w\approx -0.2$ is approximately $3\sigma$ away from $w=-1$, and $w=-0.6$ is
allowed at $1\sigma$.  Note that $h$ is not well determined, i.e. the contours
look similar for all three values of $h$, and 1$\sigma$ models exist for any
value of $h$ in the range $0.5\lsim h \lsim 0.9$.  This is not surprising, as
Figure~\ref{fig:clust_h} shows $dN/dzd\Omega$ is insensitive to the value of
$h$, with only a mild $h$--dependence through the non--power law shape of the
power spectrum.

\subsection{Expectations from the X--ray Survey}
\label{subsec:wCOSMEX}

The total number of clusters in the X--ray survey in our fiducial model is
$\approx 1000$, ten times that in the SZE survey; all X--ray clusters are
located between $0<z<1$. Figure~\ref{fig:slicex} contains expectations for the
X-ray survey; we show contours of 1, 2, and 3$\sigma$ probabilities relative to
the fiducial $\Lambda$CDM model. The qualitative features are similar to that
in the SZE case, but owing to the larger number of clusters, the constraints
are significantly stronger and the contours are narrower.  However, the
contours extend further along the $w$ axis, and the largest value of $w$
allowed at a probability better than $3\sigma$ is $w>-0.2$ (assuming that the
values of $\Omega_m$ and $h$ are not known). Although the contours are narrower
than in the SZE case, assuming that $h$ and $w$ are unknown, the allowed range
of $\Omega_m$ is similar to that in the SZE case, $0.26\lsim \Omega_m \lsim
0.42$.  Note that because of the shape and direction of the likelihood
contours, a knowledge of $h$ would not significantly improve this constraint
(although if $h$ is found to be low, then the lower limit in $\Omega_m$ would
increase).  Finally, assuming that both $h$ and $\Omega_m$ are known to high
accuracy ($\approx 3\%$), the allowed $3\sigma$ range on $w$ could be reduced
to $-1\leq w\lsim -0.85$.

\section{Results and Discussion}
\label{sec:discussion}

\subsection{Total Number vs. the Redshift Distribution}
\label{subsec:shape}

Our main results are presented in Figures~\ref{fig:slices} and
\ref{fig:slicex}, which show the probabilities of various models relative to a
fiducial $\Lambda$CDM model in the SZE and X--ray surveys.  As demonstrated by
these figures, the cluster data determine a combination of $\Omega_m$ and $w$.
In the absence of external constraints on $\Omega_{m}$ and $h$, $w$ as large as
$-0.2$ differs from $w=-1$ by $3\sigma$; while $w=-0.6$ would be $1\sigma$ away
from our fiducial $\Lambda$CDM cosmology.  Owing to the larger number of
clusters in the X--ray survey, the constrained combination of $\Omega_m$ and
$w$ is significantly narrower than in the SZE survey; the direction of the
contours is also somewhat different.  As a result, analysis of the X--ray 
survey could distinguish
a $w\approx -0.85$ model from $\Lambda$CDM at $3\sigma$ significance, provided
$\Omega_m$ is known to an accuracy of $\sim3\%$ from other studies.

It is interesting to ask whether these constraints arise mainly from the total
number of detected clusters, or from their redshift distribution.  To address
this issue, in Figure~\ref{fig:sep} we show separate likelihood contours for
the probability $P_0$ (total number of clusters, left panels), and for the
probability $P_z$ (shape of redshift distribution, right panels).  In the SZE
case, the contours of likelihood from the shape information alone are broad,
and adding these constraints to the Poisson--probability plays almost no role
in the range $w\lsim -0.7$ (the contours of $P_{\rm tot}$ and $P_0$ are very
similar). However, at larger $w$, the shape becomes increasingly
important. Adding in this information significantly reduces the allowed region
relative to the Poisson--probability alone at $w\gsim -0.7$. It is the
combination of the $P_0$ and $P_z$ contours that allows ruling out $w\gsim
-0.2$ at the $3\sigma$ level.  Note that the difference in shapes arises mostly
from the high--redshift ($z\gsim 1$) clusters (cf. Fig.~\ref{fig:clust_w}).

In the X--ray case (bottom panels in Fig.~\ref{fig:sep}), the situation is
different, because the contours of $P_0$ and $P_z$ are both much narrower.  As
a result, the contours for the combined likelihood are somewhat reduced, but
they still reach to $w\approx -0.2$ (at $\sim 2\sigma$).  Note that as in the
SZE survey, the redshift distribution (of clusters primarily in the $0<z<1$
range) plays an important role. As Figures~\ref{fig:clust_h} and
~\ref{fig:clust_om} show, the total number of clusters can be adjusted by
changing $\Omega_m$ and $h$.  In terms of the total number of clusters, $w$ is
therefore degenerate both with $\Omega_m$ and $h$: raising $w$ lowers the total
number, but this can always be offset by a change in $\Omega_m$ and/or $h$.
The bottom left panel in Fig.~\ref{fig:sep} reveals that based on $P_0$ alone,
$w=-0.2$ (and $\Omega_m=0.43$) can not be distinguished from $\Lambda$CDM even
at the $1\sigma$ level.  On the other hand, the middle panel in
Fig.~\ref{fig:slicex} shows that when the shape information is added, $w \lsim
-0.2$ follows to $2\sigma$ significance.

\myputfigure{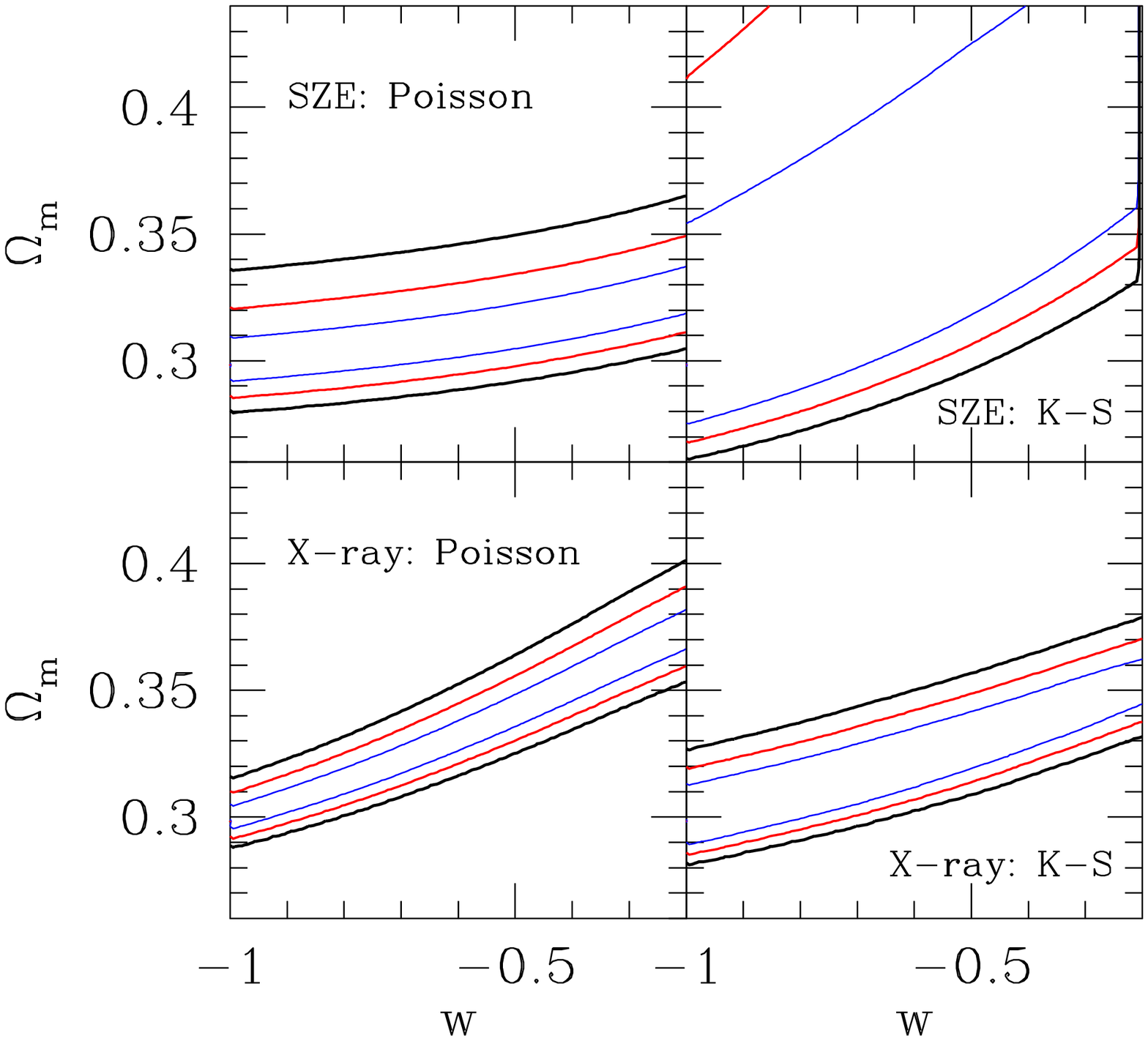}{3.35}{0.50}{-35}{-00}
\figcaption{Likelihood contours of 1, 2, and 3$\sigma$ probabilites as in
Figures \ref{fig:slices} and \ref{fig:slicex}, but when only the total number
of clusters (left panels), or only the redshift distributions (right panels)
are used to compute the likelihoods between two models.
\vspace{\baselineskip}
\label{fig:sep}}

\subsection{Discussion of Possible Systematic Uncertainties}
\label{subsec:systematics}

Our results imply that the cluster abundances in the SZE and X--ray surveys can
provide useful constraints on cosmological parameters, based on statistical
differences expected among different cosmologies.  The purpose of this section
is to summarize and quantify the various systematic uncertainties that can
affect these constraints.

\vspace{\baselineskip} {\it Knowledge of the Limiting Mass $M_{\rm min}$}.  Our
conclusions above are dependent on the chosen limiting mass, which is a
function of both redshift and cosmology.  From the discussion in
\S~\ref{subsec:parameters} we have seen that the limiting mass plays a
secondary role in the SZE survey, where the bulk of the constraint comes from
the growth function.  In comparison, we find that $M_{\rm min}$ plays an
important role in the X--ray survey.  To demonstrate the importance of
the mass limit explicitly, in Figure~\ref{fig:slicex-noM} we show the
likelihood contours in the $\Omega_m - w$ plane when the variations of the
limiting mass with cosmology are not taken into account.  Not surprisingly,
this makes the contours somewhat narrower, but nearly parallel to $w$ -- this
is consistent with our finding in Figure \ref{fig:clust_xnoM} that the mass
limit accounts for nearly all of the $w$--dependence, but it reduces the
$\Omega_m$ dependence.  Figure~\ref{fig:slicex-noM} demonstrates the need to
accurately know the limiting mass $M_{\rm min}$, and its cosmological scaling, in
the X--ray survey.

\myputfigure{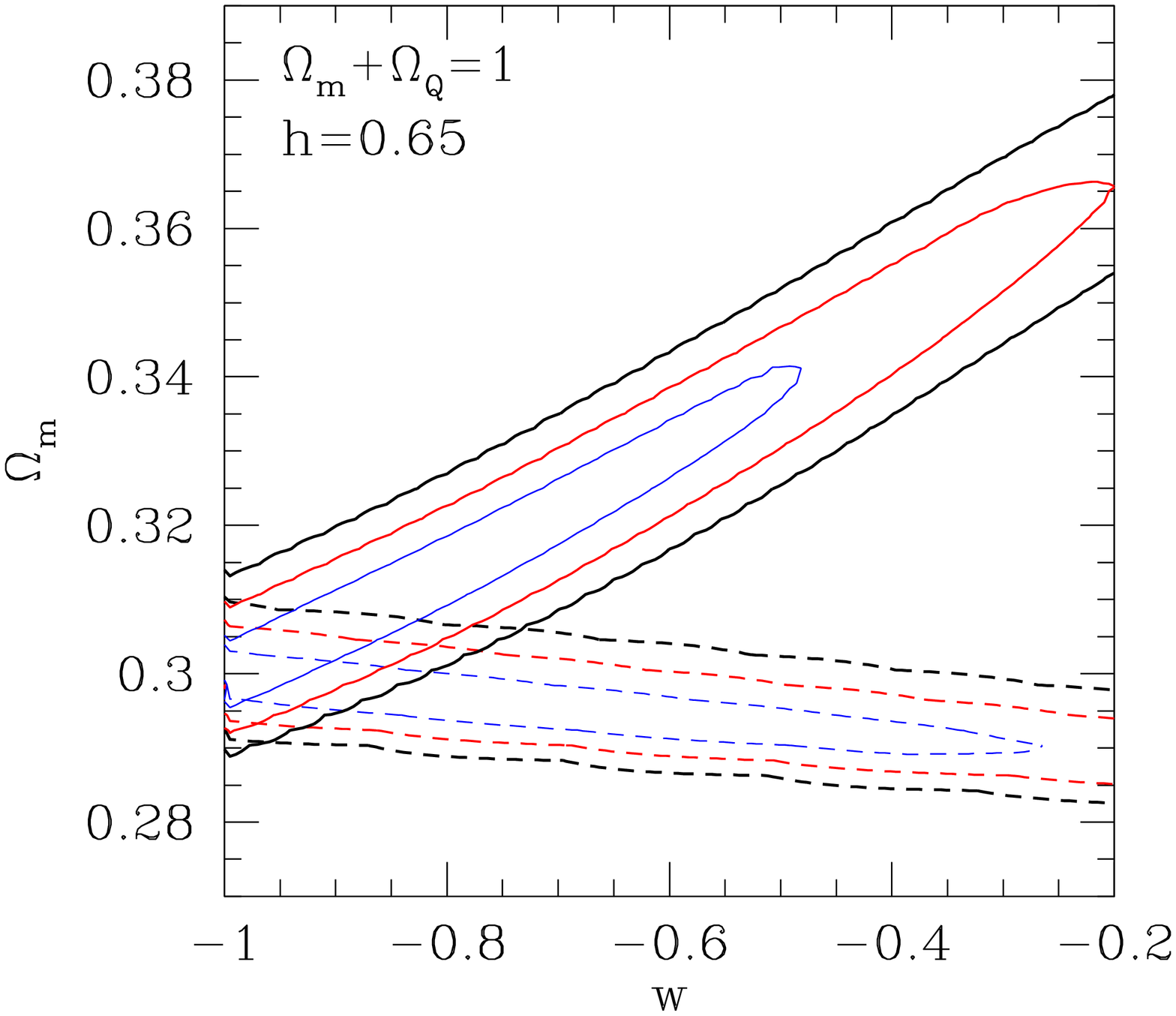}{3.2}{0.48}{-35}{-00} 
\figcaption{Likelihood contours for a fixed $h=0.65$ in the X--ray survey, as
in the middle panel of Figure~\ref{fig:slicex}, but zooming in for clarity.
The added (nearly horizontal) contours shows the allowed region when variations
of the limiting mass with cosmology are not taken into
account.
\label{fig:slicex-noM}}
\vspace{\baselineskip}

\begin{figure*}[hbt]
\vskip-0.2in
\hbox to \hsize{\hfil\hskip-0.25in\vbox to 5.6in{
\epsfysize=5.6in
\epsfbox{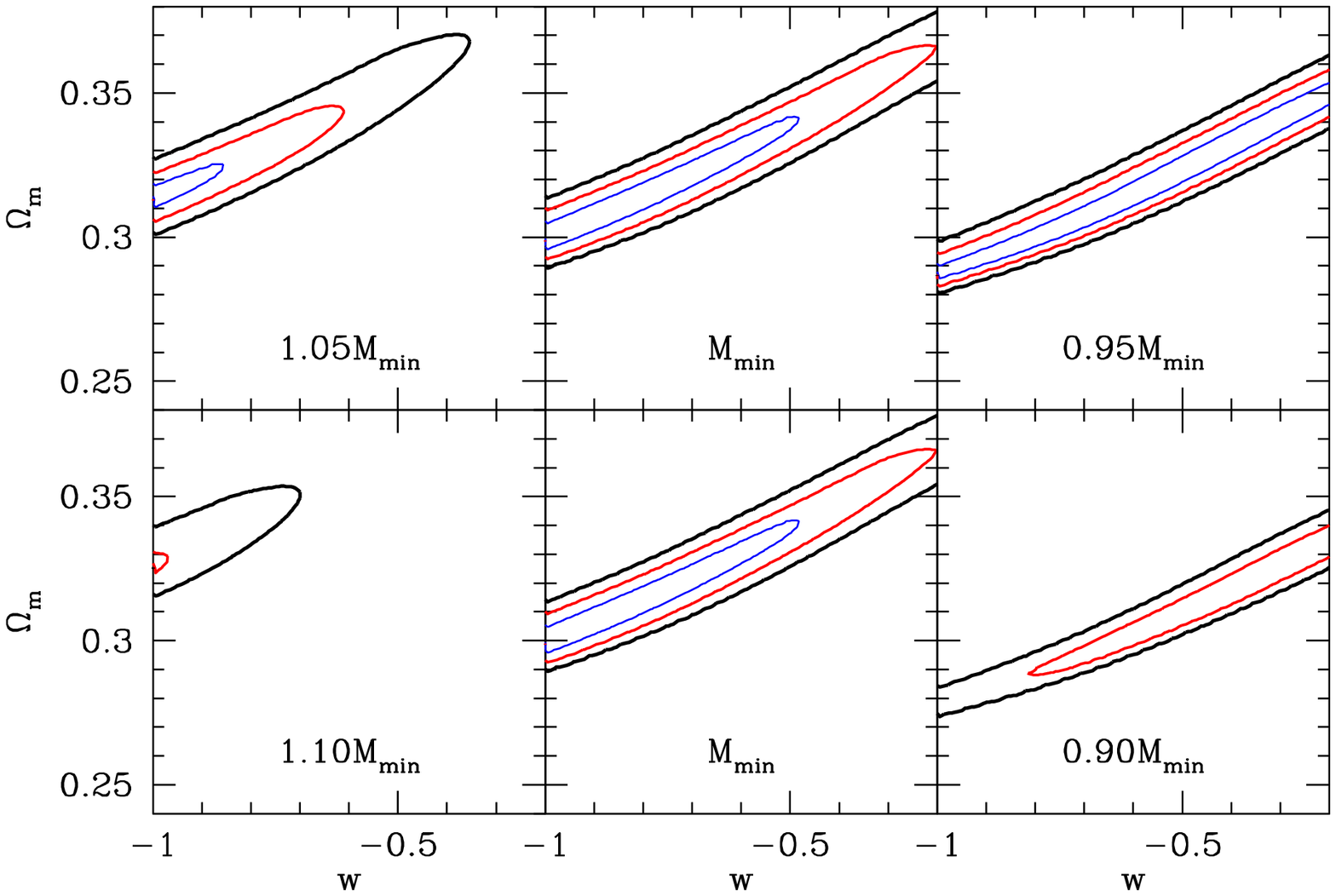}
\vfil}\hfil}\vskip-0.4in
\caption{\footnotesize The middle panels show the likelihood contours for a
fixed $h=0.65$ in the X--ray survey, as in Figure~\ref{fig:slicex}.  The upper
and lower panels show the deviations in the contours caused by either a $\pm
5\%$ or a $\pm 10\%$ offset in the $M-T$ normalization.}
\label{fig:slicex-mlim}
\vskip-0.2in
\end{figure*}

\begin{figure*}[hbt]
\vskip-0.2in
\hbox to \hsize{\hfil\hskip-0.2in\vbox to 3.3in{
\epsfysize=3.3in
\epsfbox{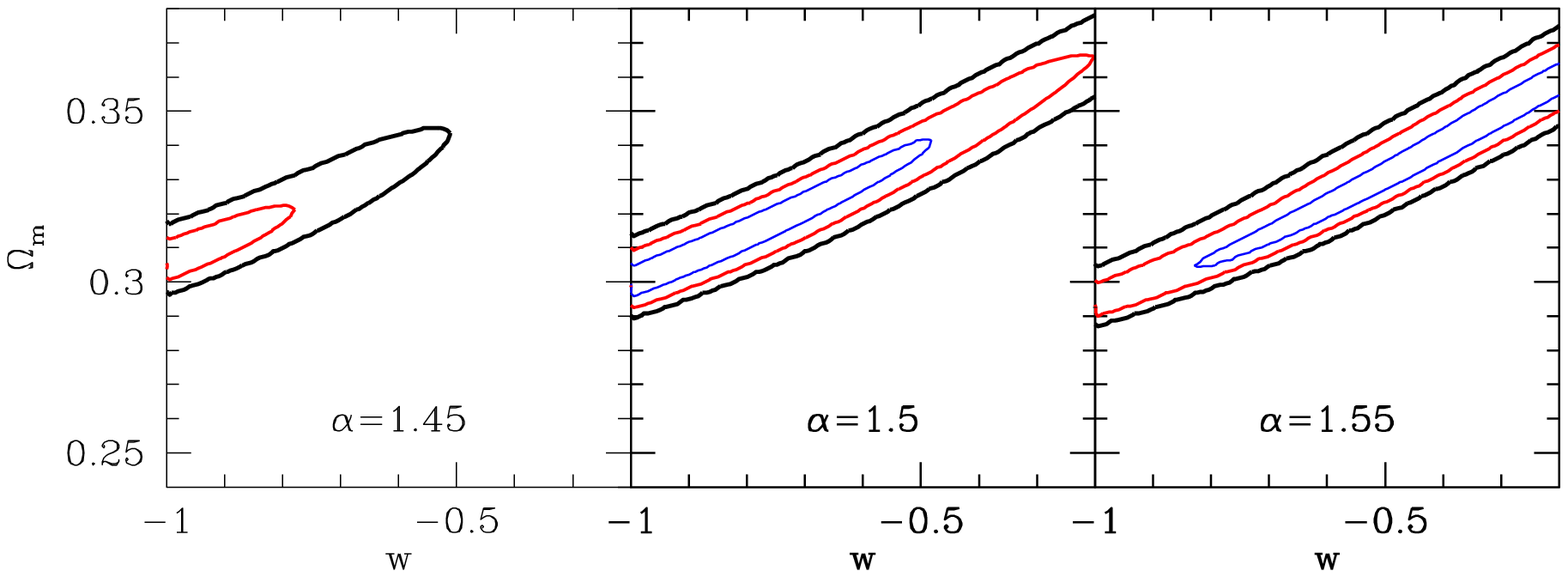}
\vfil}\hfil}\vskip-0.4in
\caption{\footnotesize The middle panels show the likelihood contours for a
fixed $h=0.65$ in the X--ray survey, as in Figure~\ref{fig:slicex}.  The other
two panels show the deviations caused by an offset in the slope $M\propto
T^\alpha$.}
\label{fig:slicex-slope}
\vskip-0.2in
\end{figure*}

Because our proposed cluster sample will have measured X--ray temperatures, the
uncertainty in our knowledge of the limiting mass will likely be dominated by
the theoretical uncertainties of the $M-T$ relation.  In order to quantify the
effect of such errors, we have performed a set of simple modifications to our
modeling of the constraints from the X--ray survey.  In all cases, we adopt the
same $M-T$ relations as we did before (cf. eq.~\ref{eq:mt}).  However, in the
fiducial model, we use a limiting mass that is altered by either $\pm 5\%$ or
$\pm 10\%$ from the mass inferred from this $M-T$ relation.  This mimics a
situation where the theoretical $M-T$ relation we apply is either 5\% or 10\%
away from the relation in the real universe.  In a second set of calculations,
we mimic a situation where the slope of the $M-T$ relation is incorrectly
modeled; i.e. we alter this slope in the fiducial model to $\alpha=1.5 \pm
0.05$.  The deviations to the likelihood contours caused by these offsets are
demonstrated in Figure~\ref{fig:slicex-mlim}, which shows the effects of the
offset in the $M-T$ normalization, and in Figure~\ref{fig:slicex-slope}, which
shows the effects of the offsets in the slope.  As the figures reveal, the
contours shift relatively little under these changes. We conclude that the
results we derive are robust, as long as we can predict the $M-T$ relation to
within $\sim 10\%$.

\begin{figure*}[b]
\vskip-0.4in
\hbox to \hsize{\hfil\hskip-0.2in\vbox to 3.3in{\epsfysize=3.3in
\epsfbox{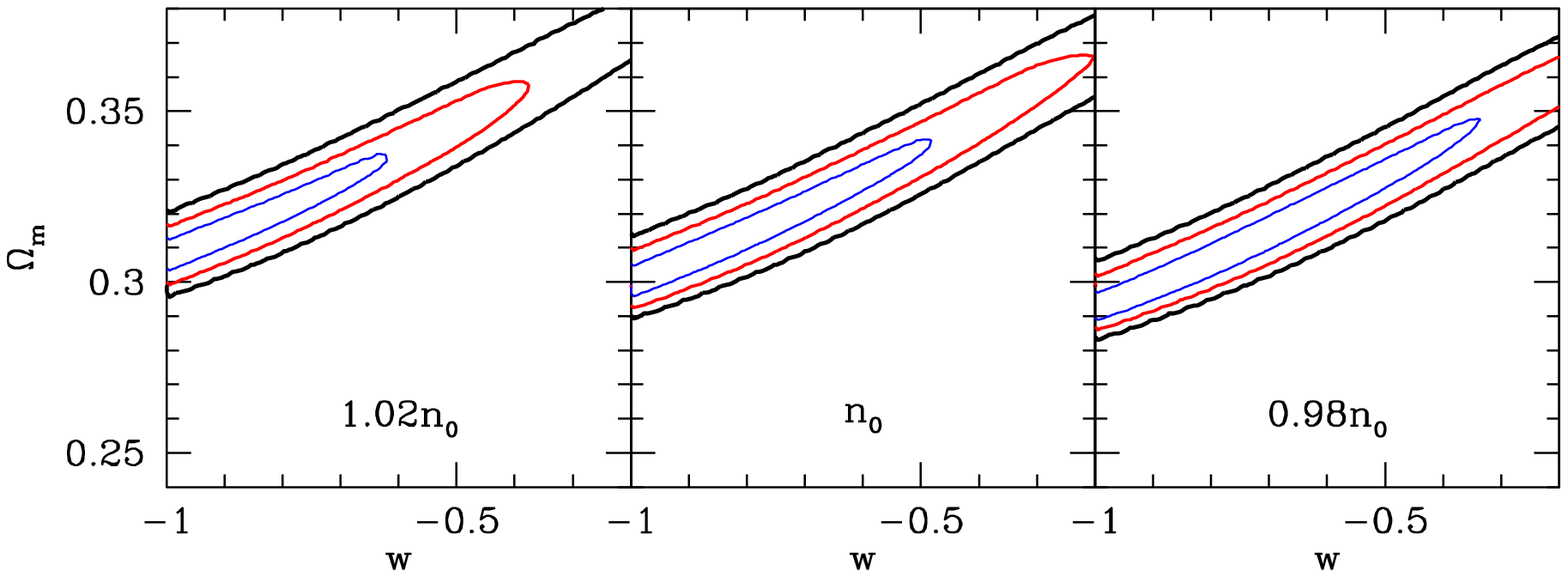} \vfil}\hfil}\vskip-0.4in
\caption{\footnotesize The middle panel shows the likelihood contours for a
fixed $h=0.65$ in the X--ray survey, as in Figure~\ref{fig:slicex}.  The left
and right panels show the deviations in the contours caused by a $\pm 2\%$
offset in the local cluster abundance determination.}
\label{fig:slicex-ncom}
\end{figure*}

In our approach, we have attempted to utilize the whole observed cluster
sample, down to the detection threshold: we had to therefore include the above
cosmological dependencies. In principle, measured cluster velocity dispersions
and X--ray temperatures (both of which are cosmology independent) could be
utilized to improve the constraints, i.e. by selecting sub--samples that
maximize the differences between models.  Further work is needed to clarify the
feasibility of this approach, as well as to quantify the accuracy to which the
dependence of $M_{\rm min}$ on $\Omega_m$, $h$, $w$, and $z$ can be predicted.

\vspace{\baselineskip} {\it Evolution of Internal Cluster Structure}.  Further
work is also required to test the cluster structural evolution models we
use. For the X--ray survey, we have assumed that the cluster
luminosity--temperature relation does not evolve, consistent with current
observations (\cite{mushotzky97}), and in the SZE survey, we have adopted the
structural evolution found in state of the art hydrodynamical simulations.
Because of the sensitivity of the survey yields to the limiting mass, cluster
structural evolution which changes the observability of high redshift clusters
can introduce systematic errors in cosmological constraints: for example, both
low $\Omega_m$ cosmologies and positive evolution of the cluster
luminosity--temperature relation increase the cluster yield in an X-ray survey.
SZE surveys are generally less sensitive to evolution than X--ray surveys,
because the X--ray luminosity is heavily dependent on the core structure (e.g.,
the presence or absence of cooling instabilities), whereas the SZE visibility
depends on the integral of the ICM pressure over the entire cluster
(Eqn.~\ref{eq:MlimSZ}).  We are testing these assertions with a new suite of
hydrodynamical simulations in scenarios where galaxy formation at high redshift
preheats the intergalactic gas before it collapses to form clusters
(\cite{bialek00}; Mohr et al. in prep).  However, most importantly, we
emphasize that because of the sensitivity of X--ray surveys to evolution, we
have only used those clusters which produce enough photons to measure an
emission weighted mean temperature.  In this case, one can directly extract the
minimum temperature $T_{lim}(z)$ of detected clusters as a function of
redshift.  Correctly interpreting such a survey requires mapping $T_{lim}(z)\to
M_{lim}(z)$ using the mass-temperature relation; the evolution of the
mass-temperature relation is less sensitive to the details of preheating than
the luminosity-temperature relation.  Thus, in a survey constructed in this
manner, it should be possible to disentangle the cosmological effects from
those caused by the evolution of cluster structure.

\vspace{\baselineskip} {\it Cluster Mass Function}.  In our treatment, we have
relied on the mass function inferred from large scale numerical simulations of
Jenkins et al. (2000).  Although we do not expect the results presented here to
change qualitatively, changes in $dN/dM$ by upto the quoted accuracy of $\sim
30\%$ could affect the exact shape of the likelihood contours shown in
Figures~\ref{fig:slices} and \ref{fig:slicex}.  It is important to test the
scaling of the mass function with cosmological parameters in future
simulations.  We have further ignored the effects of galaxy formation and
feedback on the limiting mass.  In principle, the relation between the cluster
SZE decrement and virial mass in the lowest mass clusters could be affected by
these processes.  In addition, the dependence of both the SZE decrement and the
X--ray flux likely exhibits a non--negligible intrinsic scatter.  The SZE
decrement to virial mass relation is found to have a small scatter in numerical
simulations (Metzler 1998), and to cause a negligible increase in the total
cluster yields (Holder et al. 1999).  However, the presence of scatter could
effectively lower the limiting masses in our treatment of the X--ray survey.

\vspace{\baselineskip} {\it Local Cluster Abundance}.  Perhaps the most
critical assumption is that the local cluster abundance is known to high
accuracy.  We have used this assumption to determine $\sigma_8$, i.e.  to
eliminate one free parameter -- effectively assigning ``infinite weight'' to
the cluster abundance near $z=0$.  This approach is appropriate for several
reasons.  The cosmological parameters make little difference to the cluster
abundance at $z\approx 0$, other than the volume being proportional to
$h^{-3}$.  Similarly, the study of local cluster masses is cosmologically
independent (upto a factor of $h$).  In a $10^4$ square degree survey, we find
that the total number of clusters between $0<z<0.1$, down to a limiting mass of
$2\times 10^{14}h^{-1}~{\rm M_\odot}$ is $\approx 2500$; with a random error of
only $\pm 2\%$. We have experimented with our models, assuming that the
normalization at $z=0$ is incorrectly determined by a fraction of 2\%.  In
Figure~\ref{fig:slicex-ncom}, we show the shift in the usual likelihood contour
in the X--ray survey, caused by errors in the local abundance at this level.
As the figure shows, the shift is relatively small (by about the width of the
$1\sigma$ region).  In similar calculations with errors of $\pm 4\%$, we find
shifts that are approximately twice as significant.  We conclude that for our
normalization procedure to be valid, the local cluster abundance has to be
known to an accuracy of about $\lsim 10\%$.

Although such an accuracy can be achieved by only $\sim 600$ nearby clusters
(which can be provided, for example, by an analysis of the SDSS data or perhaps
the 2MASS survey), it is interesting to consider a different approach, where
$\sigma_8$ is treated as another free parameter in addition to $\Omega_m, h,$
and $w$.  The result of such a calculation over a 4--dimensional grid is
displayed in Figure~\ref{fig:slicex-nosig}.  This figure shows the likelihood
contours along the slice $h=0.65$ through this parameter space, but in
projection along the $\sigma_8$ axis; to be compared directly with the middle
panel of Figure~\ref{fig:slicex}.  Allowing $\sigma_8$ to vary results in a
range of values $0.70 < \sigma_8 < 0.97$, and considerably expands the allowed
likelihood region.  The shape of the contours stay nearly unchanged, but their
widths along the $\Omega_m$ direction expand by approximately a factor of $\sim
4$, and their lengths along the $w$ direction increase by about a factor of 2.
We conclude that our constraints would be significantly weakened without the
local normalization (but would still be potentially useful when combined with
other data; see below).

\vspace{\baselineskip} {\it More General Cosmologies}.  In section
\ref{sec:wconstraints}, we restricted our range of models to flat CDM models.
We find that the redshift distribution of clusters in open CDM models typically
resembles that in models with high $w$.  This is demonstrated in
Figure~\ref{fig:clust_w}: both in the $w=-0.2$ and the OCDM model, the redshift
distributions are flatter and extend to higher $z$ than in $\Lambda$CDM.  We
find that OCDM models with suitably adjusted values of $\Omega_m$ and $h$ are
typically difficult to distinguish from those with $w\gsim -0.5$, but the flat
shape of $dN/dzd\Omega$ makes OCDM easily distinguishable from $\Lambda$CDM.
Note that open CDM models appear inconsistent with the recent CMB anisotropy
data from the Boomerang and Maxima experiments
(e.g. \cite{lange00,white00,bond00}).  A broader study of different
cosmological models, including those with both dark energy and curvature,
time--dependent $w$, and those with non--Gaussian initial conditions could
reveal new degeneracies, and will be studied elsewhere.

\myputfigure{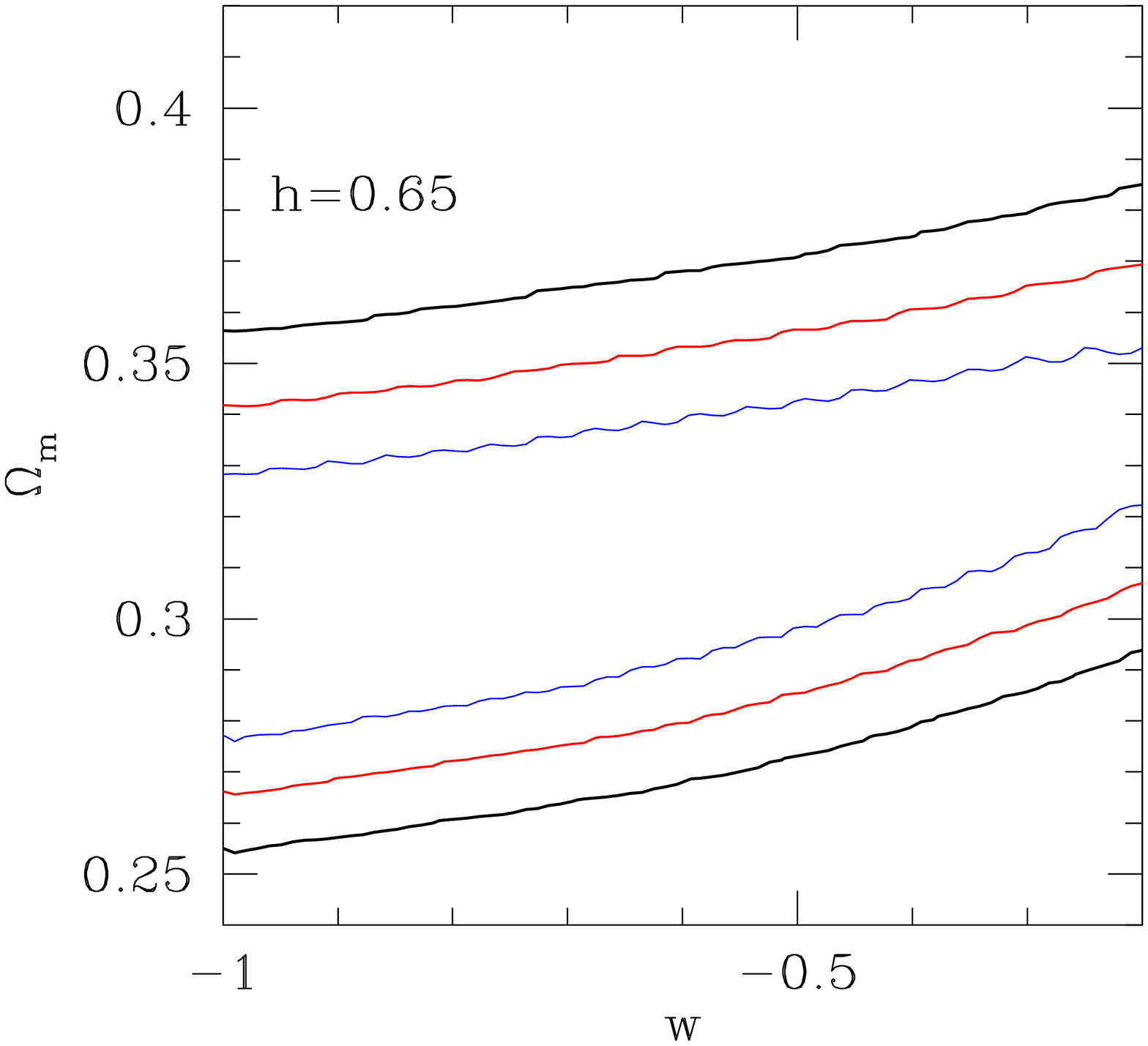}{3.25}{0.49}{-35}{-00} 
\figcaption{Likelihood contours for a fixed $h=0.65$ in the X--ray survey, as
in Figure~\ref{fig:slicex}; however, we here considered $\sigma_8$ as a free
parameter, rather than fixing its value based on the local 
abundance.  For the range of $\Omega$ and $w$ shown, models with
a likelihood better than 3$\sigma$ take on values between 
$0.7<\sigma_8<0.97$. \vspace{\baselineskip}
\label{fig:slicex-nosig}}

\subsection{Clusters versus CMB Anisotropy and High-$z$ SNe}
\label{subsec:cmb}

A useful generic feature of the likelihood contours presented here is their
difference from those expected in CMB anisotropy or Supernovae data.  Two
different cosmologies produce the same location (spherical harmonic index
$\ell_{\rm peak}$) for the first Doppler peak for the CMB temperature
anisotropy, provided they have the same comoving distance to the surface of
last scattering (cf. \cite{wang98,white98,huey99}).  Note that this is only the
most prominent constraint that can be obtained from the CMB data, with
considerable more information once the location and height of the second and
higher Doppler peaks are measured. Similarly; the apparent magnitudes of the
observed SNe constrain the luminosity distance $d_L(z)$ to $0\leq z\lsim 1$
(\cite{schmidt98,perlmutter99}).  In general, both of these types of
observations will determine a combination of cosmological parameters that is
different from the cluster constraints derived here.

\myputfigure{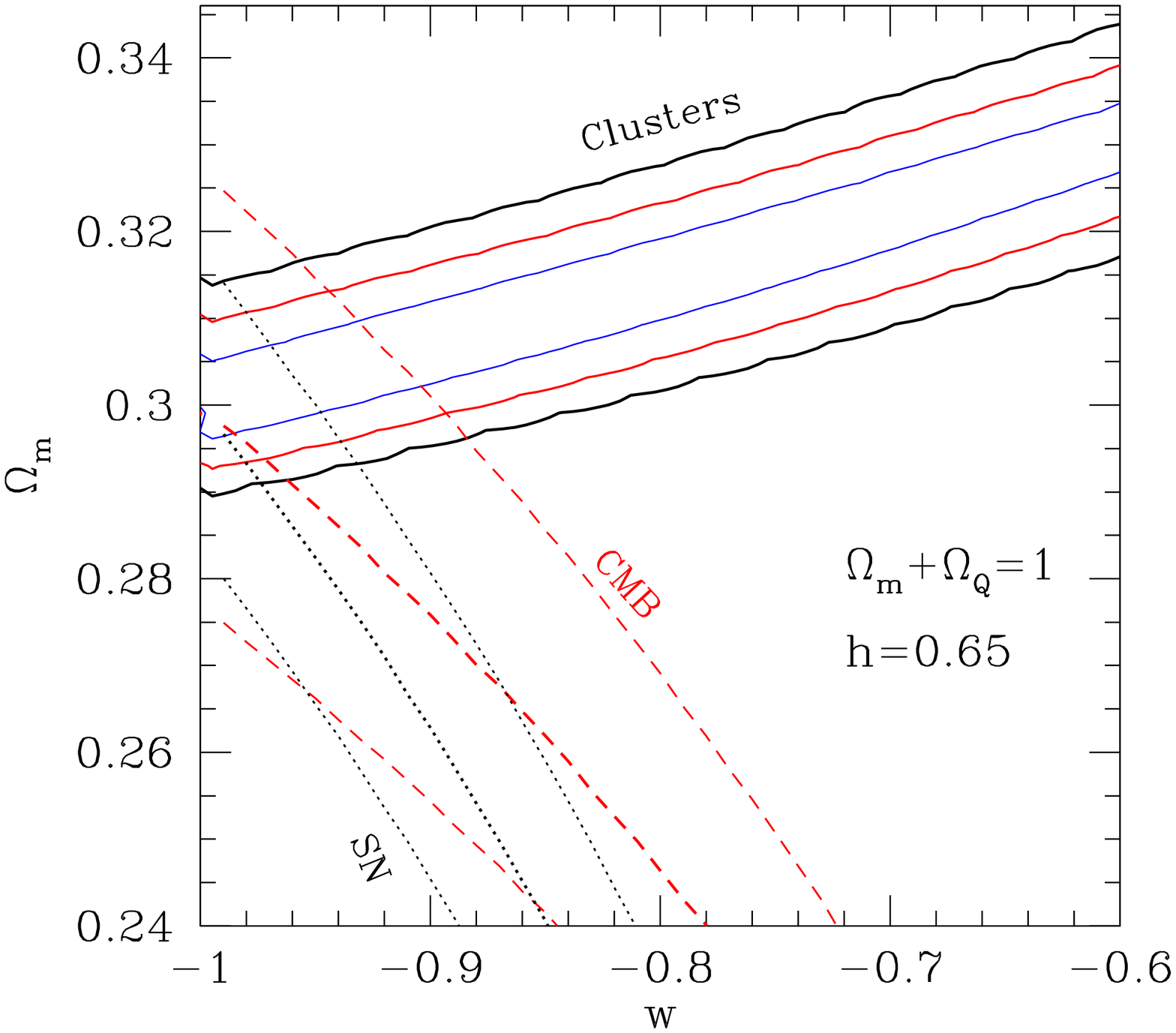}{3.25}{0.47}{-35}{-00} 
\figcaption{Likelihood contours for a fixed $h=0.65$ as in
Figure~\ref{fig:slicex}, but zooming in for clarity.  Also shown are
combinations of $w$ and $\Omega_m$ that keep the spherical harmonic index
$\ell$ of the first Doppler peak in the CMB anisotropy data constant to within
$\pm 1\%$ (dashed lines); and combinations that keep the luminosity distance to
redshift $z=1$ constant to the same accuracy.\vspace{\baselineskip}
\label{fig:slicex-zoom}}

In Figure~\ref{fig:slicex-zoom}, we zoom in on the relevant region of the
$\Omega_m-w$ plane in the X--ray survey, and compare the cluster constraints to
those expected from CMB anisotropy or high--$z$ SNe.  The three dashed curves
correspond to the CMB constraints: the middle curve shows a combination of
$\Omega_m$ and $w$ that produces the constant $\ell_{\rm peak}\approx 243$
obtained in our fiducial $\Lambda$CDM model (using the fitting formulae from
White 1998 for the physical scale $k_{\rm peak}$); the other two dotted curves
bracket a $\pm 1\%$ range around this value.  Similarly, the dotted curves
correspond to the constraints from SNe.  The middle curve shows a line of
constant $d_L$ at $z=1$ that agrees with the $\Lambda$CDM model; the two other
curves produce a $d_L$ that differs from the fiducial value by $\pm 1\%$.  As
the figures show, the lines of CMB and SNe parameter degeneracies run somewhat
unfavorably parallel to each other; however, both of those degeneracies are
much more complementary to the direction of the parameter degeneracy in cluster
abundance studies.  In particular, the maximum allowed value of $w$, using both
the CMB or SNe data, is $w\approx -0.8$; while this is reduced to $w\approx
-0.95$ when the cluster constraints are added.  Note that in
Figure~\ref{fig:slicex-zoom}, we have assumed a fixed value of $h=0.65$;
however, we find that relaxing this assumption does not significantly change
the above conclusion.  The CMB and SNe constraints depend more sensitively on
$h$ than the cluster constraints do: as a result, the confidence regions do not
overlap significantly even in the three--dimensional ($w,\Omega_m,h$) space.

The high complementarity of the cluster constraint to those from the other two
methods can be understood based on the discussions in
$\S$~\ref{subsec:parameters}.  To remain consistent with the CMB and SNe Ia
constraints, an increase in $w$ must be coupled with a decrease in
$\Omega_{m}$; however, both increasing $w$ and lowering $\Omega_m$ raises the
number of detected clusters.  To keep the total number of clusters constant, an
increase in $w$ must be balanced by an increase in $\Omega_m$.  Note that this
statement is true both for the SZE and the X--ray surveys. Combining the
cluster constraints with the CMB and SNe Ia constraints will therefore likely
result in improved estimates of the cosmological parameters, and we do not
expect this conclusion to rely on the details of the two surveys considered
here.

\section{Conclusions}
\label{sec:conclusions}

We studied the expected evolution of galaxy cluster abundance from $0\lsim z
\lsim 3$ in different cosmologies, including the effects of variations in the
cosmic equation of state parameter $w\equiv p/\rho$.  By considering a range of
cosmological models, we quantified the accuracy to which $\Omega_m$, $w$, and
$h$ can be determined in the future, using a 12~deg$^{2}$ Sunyaev-Zel'dovich
Effect survey and a deep 10$^{4}$~deg$^{2}$ X-ray survey.  In our analysis, we
have assumed that the local cluster abundance is known accurately: we find that
in practice, an accuracy of $\sim 5\%$ is sufficient for our results to be
valid.

We find that raising $w$ significantly flattens the redshift--distribution,
which can not be mimicked by variations in either $\Omega_m$, $h$, which affect
essentially only the normalization of the redshift distribution. As a result,
both surveys will be able to improve present constraints on $w$.  In the
$\Omega_m-w$ plane, both the SZE and X--ray surveys yield constraints that are
highly complementary to those obtained from the CMB anisotropy and high--$z$
SNe.  Note that the SZE and X--ray surveys are themselves somewhat
complementary. In combination with these data, the SZE survey can determine
both $w$ and $\Omega_m$ to an accuracy of $\approx 10\%$ at $3\sigma$
significance.  Further improvements will be possible from the X--ray survey.
The large number of clusters further alleviates the degeneracy between $w$ and
both $\Omega_m$ and $h$, and, as a result, the X--ray sample can determine $w$
to $\approx 10\%$ and $\Omega_m$ to $\approx 5\%$ accuracy, in combination with
either the CMB or the SN data.

Our work focuses primarily on the statistics of cluster surveys.  We have
provided an estimate of the scale of various systematic uncertainties.  Further
work is needed to clarify the role of these uncertainties, arising especially
from the analytic estimates of the scaling of the mass limits with cosmology,
the dependence of the cluster mass function on cosmology, and our neglect of
issues such as galaxy formation in the lowest mass clusters.  However, our
findings suggest that, in a flat universe, the cluster data lead to tight
constraints on a combination of $\Omega_m$ and $w$, especially valuable because
of their high complementarity to those obtained from the CMB anisotropy or
Hubble diagrams using SNe as standard candles.

\acknowledgements

We thank L. Hui for useful discussions, D. Eisenstein, M. Turner, D. Spergel
and the anonymous referee for useful comments, and J. Carlstrom and the COSMEX
team for providing access to instrument characteristics required to estimate
the yields from their planned surveys.  ZH is supported by the DOE and the NASA
grant NAG 5-7092 at Fermilab, and by NASA through the Hubble Fellowship grant
HF-01119.01-99A, awarded by the Space Telescope Science Institute, which is
operated by the Association of Universities for Research in Astronomy, Inc.,
for NASA under contract NAS 5-26555.  JJM is supported by Chandra Fellowship
grant PF8-1003, awarded through the Chandra Science Center.  The Chandra
Science Center is operated by the Smithsonian Astrophysical Observatory for
NASA under contract NAS8-39073.

\end{document}